\documentclass{aa}
\usepackage{graphicx}
\usepackage{xspace} 
\usepackage{caption}
\usepackage{subcaption}
\usepackage{multirow}
\usepackage{dcolumn}
\newcolumntype{d}[1]{D{.}{.}{#1}}
\usepackage{hyperref}
\usepackage{xcolor}
\usepackage{soul}
\usepackage{ulem}
\hypersetup{
    colorlinks,
    linkcolor={red!50!black},
    citecolor={blue!50!black},
    urlcolor={blue!80!black},
    draft,
    hidelinks
}
\usepackage{txfonts}
\usepackage{amsmath}
\usepackage{natbib}
\bibpunct{(}{)}{;}{a}{}{,} 
\usepackage{lscape}
\usepackage{multicol}
\usepackage[version=4]{mhchem}
\usepackage{booktabs}
\usepackage{xcolor}
\newcommand{\meng}[1]{#1}
\newcommand{\modc}[1]{#1}
\newcommand{\modb}[1]{\iffalse #1 \fi}
\newcommand{\moda}[1]{#1}

\usepackage[detect-all,load=abbr)]{siunitx}
\newcommand{\hii}{H{\sc ii }}

\newcommand{\sgr}{Sgr\,B2 }

\begin{document} 

\title{The physical and chemical structure of Sagittarius B2}
\titlerunning{Non-thermal emission in the envelope of Sgr\,B2}
\subtitle{V. Non-thermal emission in the envelope of Sgr\,B2}
\author{F.~Meng\inst{1},
 {\'A}.~S{\'a}nchez-Monge\inst{1},
 P.~Schilke\inst{1},
 M.~Padovani\inst{2},
 A. Marcowith\inst{3},
 A.~Ginsburg\inst{4},
 A.~Schmiedeke\inst{5},
 A.~Schw{\"o}rer\inst{1},
 C.~DePree\inst{6},
 V.~S.~Veena\inst{1},
\and
Th.~M{\"o}ller\inst{1}
          }
\authorrunning{F. Meng et al.}
\institute{I.\ Physikalisches Institut, Universit\"at zu K\"oln, Z\"ulpicher Str.\ 77, D-50937 K\"oln, Germany\\
           \email{meng@ph1.uni-koeln.de}
           \and
           INAF–Osservatorio Astrofisico di Arcetri, Largo E. Fermi 5, 50125 Firenze, Italy
           \and
           Laboratoire Univers et Particules de Montpellier, UMR 5299 du CNRS, Universit{\'e} de Montpellier, place E. Bataillon, cc072, 34095 Montpellier, France
           \and
           Jansky Fellow of the National Radio Astronomy Observatory, 1003 Lopezville Rd., Socorro, NM 87801, USA
           \and
           Max Planck Institute for Extraterrestrial Physics, Giessenbachstrasse 1, D-85748 Garching, Germany
           \and
           Agnes Scott College, 141 E. College Ave., Decatur, GA 30030, USA
           }

\date{Received ; accepted }
 
\abstract{The giant molecular cloud Sagittarius B2 (hereafter Sgr\,B2) is
the most massive region with ongoing high-mass star formation in the Galaxy. In the southern region of the 40-pc large envelope of Sgr\,B2, we encounter the Sgr\,B2(DS) region, which hosts more than 60 high-mass protostellar cores distributed in an arc shape around an extended \hii region. Hints of non-thermal emission have been found in the \hii region associated with Sgr\,B2(DS).} 
{We seek to characterize the spatial structure and the spectral energy distribution of the radio continuum emission in Sgr\,B2(DS). We aim to disentangle the contribution from the thermal and non-thermal radiation, as well as to study the origin of the non-thermal radiation.} 
{We used the Very Large Array in its CnB and D configurations, and in the frequency bands C (4--8~GHz) and X (8--12~GHz) to observe the whole Sgr\,B2 complex. Continuum and radio recombination line maps are obtained.} 
{We detect radio continuum emission in Sgr\,B2(DS) in a bubble-shaped structure. From 4 to 12~GHz, we derive a spectral index between $-1.2$ and $-0.4$,  indicating the presence of non-thermal emission. We decomposed the contribution from thermal and non-thermal emission, and find that the thermal component is clumpy and more concentrated, while the non-thermal component is more extended and diffuse. The radio recombination lines in the region are found to be not in local thermodynamic equilibrium  but stimulated by the non-thermal emission.} 
{Sgr\,B2(DS) shows a mixture of thermal and non-thermal emission at radio wavelengths. The thermal free-free emission is likely tracing an \hii region ionized by an O7 star, while the non-thermal emission can be generated by relativistic electrons created through first-order Fermi acceleration. We have developed a simple model of the Sgr\,B2(DS) region and found that first-order Fermi acceleration can reproduce the observed flux density and spectral index.}


\maketitle

\section{Introduction}\label{s:intro}

  The giant molecular cloud Sagittarius B2 (hereafter Sgr\,B2) is the most massive  ($\sim 10^7\ M_{\odot}$, see e.g., \citealt{Goldsmith:1990aa}) region with ongoing high-mass star formation in the Galaxy, and has a higher density ($>10^5\rm\ cm^{-3}$) and dust temperature ($\sim$50--70~K) compared to other star forming regions in the Galactic plane \citep[see e.g.,][]{Ginsburg:2016aa, Schmiedeke:2016aa,Sanchez-Monge:2017aa}. Sgr\,B2 is located at a distance of $8.34 \pm 0.16$~kpc, at only 100~pc in projection from the Galactic center \citep{Reid:2014aa}\footnote{
    A new distance to the Galactic center has been measured to be $8.127 \pm 0.031$~kpc \citep{Gravity-Collaboration:2018aa}. For consistency with the paper published within the same series of studies of Sgr\,B2, we use the distance reported by \citet{Reid:2014aa}. }.
  In the central $\sim2\ \rm pc$, there are the two well-known \modb{and studied} hot cores Sgr\,B2(N) and Sgr\,B2(M) \citep[see e.g.,][]{Schmiedeke:2016aa,Sanchez-Monge:2017aa}, which contain at least 70 high-mass stars \citep[see e.g.,][]{Gaume:1995aa,De-Pree:1998aa,De-Pree:2014aa}. \moda{A larger envelope} surrounding the two hot cores  has a radius of 20~pc and contains more than 99\% of the total mass of Sgr\,B2 \citep[][see also their Fig.~1 for a \moda{sketch} of this region]{Schmiedeke:2016aa}. The relatively high temperature\moda{s, densities}, and pressure, as well as its proximity to the Galactic center make Sgr\,B2 a good target for studying high-mass star formation in extreme environments. With this in mind, we have started a project to characterize the physical and chemical properties via observations and modeling \citep{Schmiedeke:2016aa,Sanchez-Monge:2017aa,Pols:2018aa,Schworer:2019aa}. \moda{The current work is the fifth paper in the series.} While the first observational studies focused on the two central hot \modb{molecular} cores Sgr\,B2(N) and Sgr\,B2(M), in this work we investigate the \moda{physical properties of the} envelope they are embedded in.

  Previous studies of Sgr\,B2 suggest the presence of star forming activities throughout the envelope and not only in the central regions Sgr\,B2(N) and (M). \citet{Martin-Pintado:1999aa} report the presence of filament-, \moda{arc-} and shell-shaped dense gas likely produced by stellar feedback. Recent observations with ALMA at 3~mm revealed more than 200 high-mass protostellar cores distributed throughout the envelope \citep{Ginsburg:2018aa}. A remarkable feature is the presence of about 60 dense cores in the southern region, following an arc-shape distribution. This \moda{group of cores} is spatially coincident with the arc-shaped feature detected in \ce{NH3} by \citet{Martin-Pintado:1999aa}, which is referred to as Deep South (hereafter DS) by \citet{Ginsburg:2018aa}. At the position of Sgr\,B2(DS), \citet{Mehringer:1993aa} report the detection of radio continuum and radio recombination line (RRL) emission \citep[see also][]{LaRosa:2000aa,Law:2008ab}. Such continuum emission at radio wavelengths \moda{along} with RRLs is usually related to thermal ionized gas from \hii regions \citep[e.g.,][]{Kurtz:2002aa,Kurtz:2005aa}. However, \moda{along with the thermal emission detected in previous studies}, non-thermal emission,  has also been found in Sgr\,B2 due to relativistic electrons \moda{\citep{LaRosa:2005aa,Hollis:2007aa,Protheroe:2008aa,Jones:2011aa}}. \moda{\citet{Yusef-Zadeh:2007aa,Yusef-Zadeh:2013aa,Yusef-Zadeh:2016aa} studied the presence of non-thermal emission in the Galactic Center region including Sgr\,B2. Particularly, observations at 255~MHz, 327~MHz, and 1.4~GHz \citep{Yusef-Zadeh:2007ab} revealed non-thermal emission in Sgr\,B2(M).} The  thermal and non-thermal contribution to the radio continuum emission can be distinguished by their spectral energy distribution (hereafter SED). The SEDs of both thermal and non-thermal emission can be described by power-laws $S_{\nu}\propto \nu^{\alpha}$, where $\alpha$ is the so-called spectral index, which varies from $-0.1$ to $+2$ for thermal emission \citep[see e.g.,][]{Sanchez-Monge:2013ab} and becomes significantly negative for non-thermal emission \citep[e.g., $\alpha = -0.8$, see][]{Platania:1998aa}.

  In this paper, we present Very Large Array (VLA) observations of the whole Sgr\,B2 region in the frequency regime 4--12~GHz, \moda{with configurations BnC and D. Thus, this study focuses on spatial scales from $0.2$--$10$~pc.} 
  The higher sensitivity and resolution compared to previous studies allow us to study in detail the presence and properties of radio-continuum sources, both thermal and non-thermal, throughout Sgr\,B2. Although the observations cover the whole envelope of Sgr\,B2, we pay special attention to Sgr\,B2(DS). In \moda{Sect.}~\ref{s:obs} we describe the observations as well as the data reduction process. \moda{Section}~\ref{sec:results} shows the results, and the study of the spectral index. \moda{In Sect.~4 we aim at decomposing the thermal and non-thermal components dominating the emission at radio wavelengths, while in Sect.~5 we discuss on the origin of the non-thermal emission. Finally, we summarize our study in Sect.~6.}

\section{Observations and data reduction}\label{s:obs}

  \begin{table}
    \caption{Observed and stacked RRLs.}
    \label{t:rrls}

    \centering{
    \begin{tabular}{lcc}
    \toprule
    \hline \noalign{\smallskip}
    RRL & $\nu_0$ (MHz) & Stacked ($\nu$)\tablefootmark{a}\\
    \hline \noalign{\smallskip}
       H115$\alpha$ & \phantom{0}4268.14 &    \multirow{4}{*}{RRL 4.4 GHz} \\                            
       H114$\alpha$ & \phantom{0}4380.95 &                                     \\         
       H113$\alpha$ & \phantom{0}4497.78 &                                     \\         
       H112$\alpha$ & \phantom{0}4618.79 &                                     \\     \hline \noalign{\smallskip}
       H100$\alpha$ & \phantom{0}6478.76 &    \multirow{4}{*}{RRL 6.8 GHz}\\        
       H99$\alpha$  & \phantom{0}6676.08 &                                  \\                           
       H98$\alpha$  & \phantom{0}6881.49 &                                 \\        
       H97$\alpha$  & \phantom{0}7095.41 &                                 \\         \hline \noalign{\smallskip}
       H92$\alpha$  & \phantom{0}8309.38 &    \multirow{5}{*}{RRL 8.9 GHz} \\        
       H91$\alpha$  & \phantom{0}8584.82 &                                  \\        
       H90$\alpha$  & \phantom{0}8872.57 &                                  \\                           
       H89$\alpha$  & \phantom{0}9173.32 &                                  \\        
       H88$\alpha$  & \phantom{0}9487.82 &                                  \\        \hline \noalign{\smallskip}
       H87$\alpha$  & \phantom{0}9816.86 &    \multirow{5}{*}{RRL 10.5 GHz} \\        
       H86$\alpha$  & 10161.30 &                                   \\        
       H85$\alpha$  & 10522.04 &                                   \\        
       H84$\alpha$  & 10900.06 &                                   \\                          
       H83$\alpha$  & 11296.41 &                                   \\        
    \bottomrule
              
      \end{tabular}
      }
      \tablefoot{
\tablefoottext{a}{\moda{
The frequencies of the stacked RRLs are labels corresponding to the average frequency of the stacked lines. These values do not correspond to actual transition frequencies, and are only used in the excitation analysis of Sect.~\ref{subsec:stimulated_rrls}.}
}
}
    \end{table}

  We \moda{used} the VLA in \moda{its} CnB and D configurations to observe the entire Sgr\,B2 complex in the frequency bands C (4--8~GHz) and X (8--12~GHz). The observations with the CnB configuration were conducted on May 3 and 5, 2016 (project 16A-195). On February 22 and 23, 2017, the D configuration was employed for the observations (project 17B-063). The continuum emission was observed by combining a total of 64 spectral windows with a bandwidth of 128~MHz each. Alongside the continuum observations, high-resolution spectral windows were used for line observations. Eighteen RRLs in the frequency range from 4 to 12~GHz were observed with a spectral resolution between 31.25 and 62.5~kHz (1--2~$\mathrm{km\ s^{-1}}$). In Table~\ref{t:rrls}, we list the rest frequencies of the 18 RRLs. We used the mosaic mode to cover the whole extent of Sgr\,B2 ($\sim 20^{\prime}\times20^{\prime}$). A total of 10 and 18 pointings, with primary beam sizes of 7.5\arcmin\ and 4.5\arcmin\ for the C and X bands, respectively, were used. The quasar 3C286 was used as a flux and bandpass calibrator. The SED of 3C286 from 0.5 to 50~GHz was measured by \citet{Perley:2013aa}, with a flux of $5.059\pm0.021$~Jy at 8.435~GHz and a spectral index of $-0.46$. The quasar J1820-2528, the flux of which is 1.3~Jy in the C and X bands, was used as phase calibrator. The calibration was done using the standard VLA pipelines provided by the NRAO\footnote{
    The National Radio Astronomy Observatory is a facility of the National Science Foundation operated under cooperative agreement by Associated Universities, Inc.}.

  Calibration and imaging were done in Common Astronomy Software Applications (CASA) 4.7.2 \citep{McMullin:2007aa}. The calibrated measurement sets of the CnB and D array data were concatenated to improve the \textit{uv} sampling. Self-calibration was performed to reduce imaging artifacts and improve the final sensitivity of the image. Three loops of phase-only self-calibration were conducted, with solution time intervals of two seconds, equal to the integration time of the observations. The self-calibrated measurement sets were used for imaging. \moda{All the pointings of the mosaic in each band were primary beam corrected and the mosaic was imaged using the CASA task \texttt{tclean}.} With a robust factor of 0.5, the images of the C and X bands have \moda{synthesized beams} of 2.7\arcsec$\times$2.5\arcsec, with a position angle ${\rm (PA)}$ of $-82^\circ$ and 1.8\arcsec$\times$1.5\arcsec (${\rm PA}=76^\circ$), respectively. The PA is defined positive north to east. Under such resolutions, the root mean square (RMS) noise of the C and X band images are 0.4~mJy/beam and 0.2~mJy/beam respectively.  \moda{This RMS noise is limited by dynamic range effects due to the bright emission in the region. We achieve a dynamic range of about 4000 and 2000 for the X- and C-band images, respectively.} \modb{The Sgr\,B2(DS) region of the X and C-band images has dynamic range of 60 and 80, respectively.} To investigate the SED over the whole 4--12~GHz range, the measurement sets of the C and X bands were divided into 12 frequency ranges. Thus, 12 images from 4 to 12~GHz were obtained. The \textit{uv} \moda{coverage was} restricted to 0.6--50~k$\lambda$ for each frequency range \moda{to ensure that every image is sensitive to same spatial scales}. The 12 images are convolved to a final circular Gaussian beam of 4\arcsec. In addition to the continuum emission, we also imaged the 18 RRLs. In order to increase the signal-to-noise ratio, we stacked the neighboring RRLs to produce a final set of four stacked RRLs (see Table~\ref{t:rrls} for details). \moda{All the RRL images were resampled to a common spectral resolution of 2.5~km~s$^{-1}$.}  Finally, and with the aim of compensating for the low sensitivity (3~mJy/beam per 2.5-km/s channel), the images of the four stacked RRLs are smoothed to a resolution of 8\arcsec, resulting in a final sensitivity of 1~mJy/beam per 2.5-km/s channel.

  \begin{figure*}[ht]
    \begin{center}
    \begin{tabular}{c}
            \includegraphics[width=0.99\textwidth]{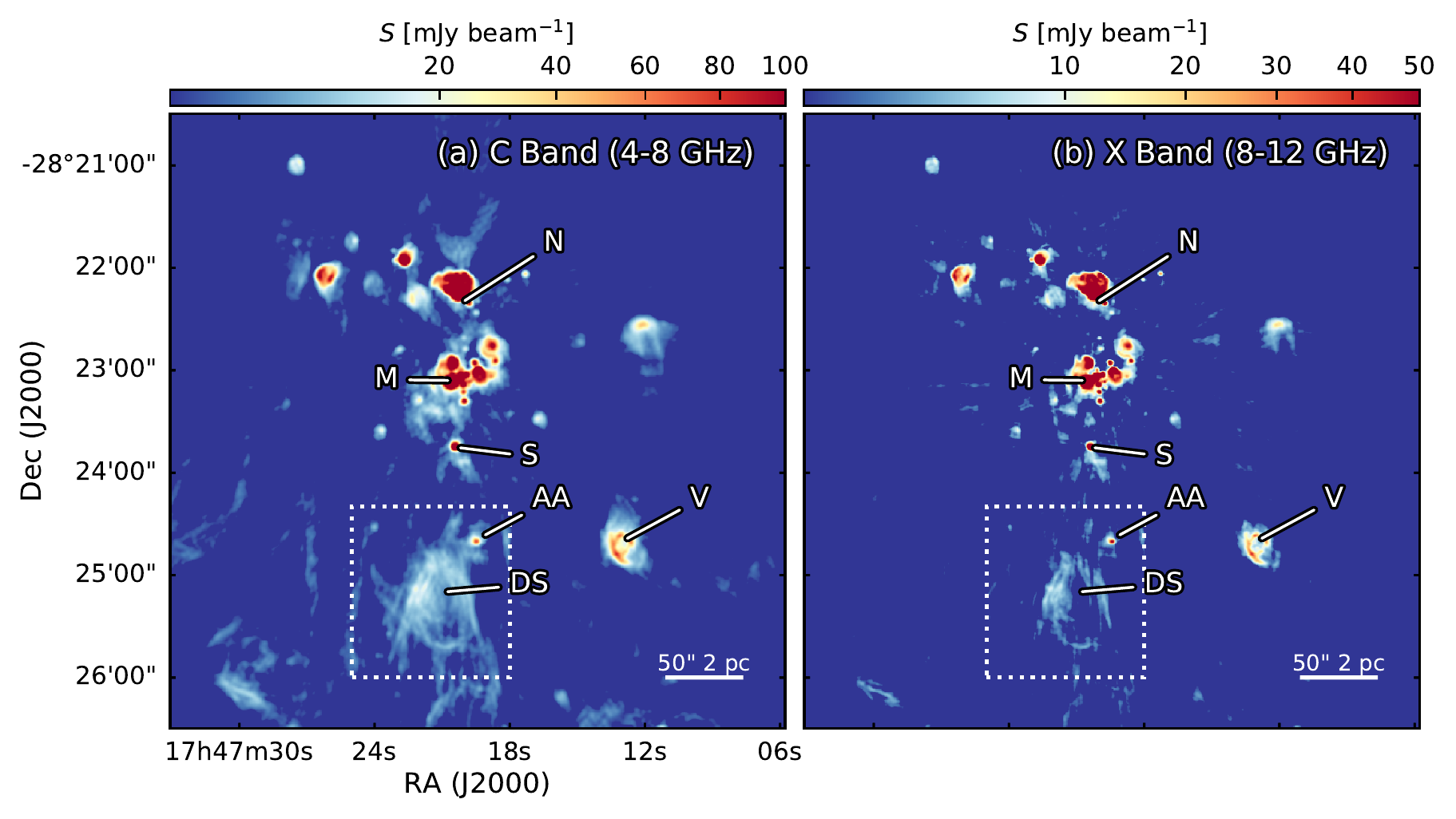} \\
    \end{tabular}
    \caption{\moda{Continuum images} of \sgr in C (panel a) and X (panel b) bands. \moda{Relevant regions are marked with their names \citep[see][]{Mehringer:1993aa}.} The dashed boxes mark the region of DS. The \moda{synthesized beam} of \moda{the} C and X \moda{images} are 2.7\arcsec$\times$2.5\arcsec and 1.82\arcsec$\times$1.53\arcsec\moda{,} respectively}
    \label{f:c-x-band}
    \end{center}
  \end{figure*}

\section{Results} 
  \label{sec:results}

  In this section~we discuss the distribution of the radio continuum emission in the Sgr\,B2 region, characterize the spectral index, and study the RRL emission. We pay special attention to the Sgr\,B2(DS) region.

  \begin{figure}[h]
    \centering
    \begin{tabular}{c}
            \includegraphics[width=0.49\textwidth]{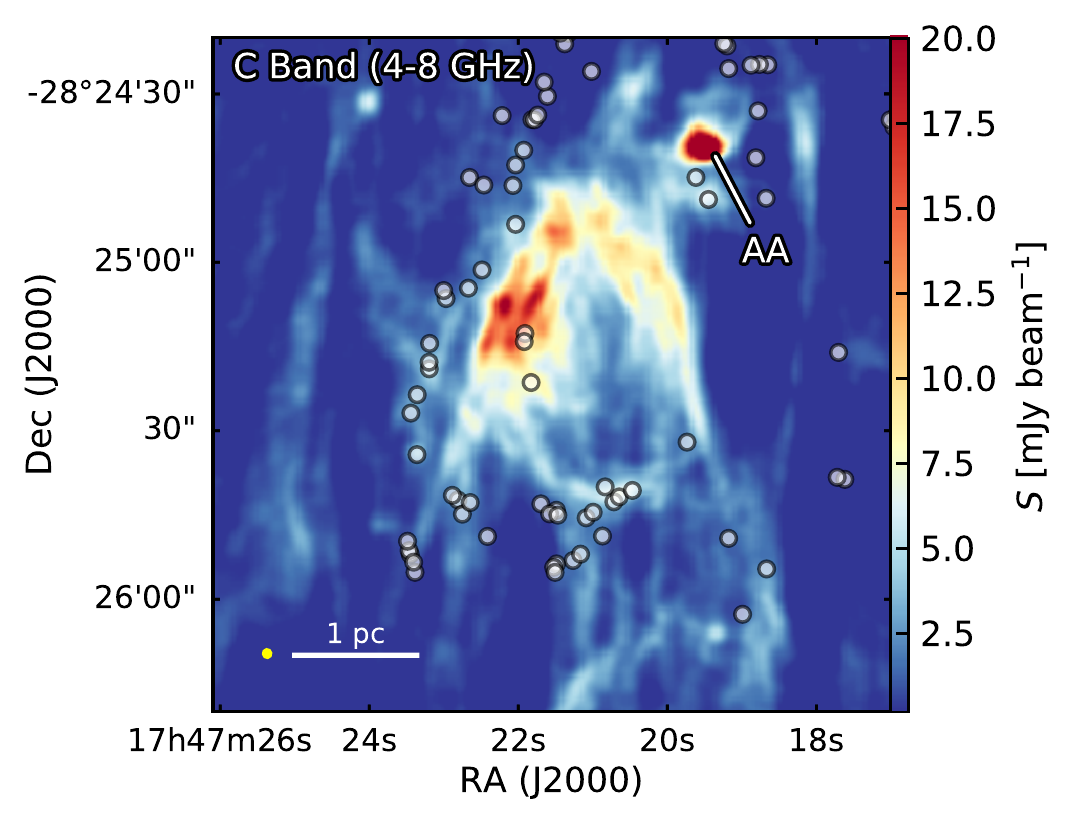}
    \end{tabular}
    \caption{\moda{C band (4--8~GHz) continuum emission map of Sgr\,B2(DS).} The circles \moda{mark} the positions of the high-mass protostellar cores identified by \citet{Ginsburg:2018aa}. \moda{The synthesized beam} is shown as \moda{a} yellow ellipse at the bottom left corner.}
    \label{f:almacoresonvla}
  \end{figure}

  \subsection{Ionized gas in Sgr\,B2(DS)}

  The images of the Sgr\,B2 region in the C and X bands are shown in Fig.~\ref{f:c-x-band}. Some relevant objects such as the regions Sgr\,B2(N), (M) and (S) are indicated, as well as \moda{source} V and the H{\sc ii} region AA \citep[see][]{Mehringer:1993aa}. The DS region is the main focus of this article. It appears as a bubble-like structure with an outer-diameter of $\sim$1.5~pc. The thickness of the bubble edge is $\sim$0.3~pc, with the eastern part being stronger than the western region of the bubble, in both X and C-band images. Filaments and \moda{arcs} are found along the edge of the bubble, while the emission towards the center is just at the RMS noise level.

  The radio continuum emission in DS is spatially connected to the high-mass dust cores revealed by previous ALMA observations at millimeter wavelengths \citep{Ginsburg:2018aa}. As shown in Fig.~\ref{f:almacoresonvla}, more than 60 dust cores are located at the outskirts of the bubble-like structure, with a larger population of dense cores to the east.

  \subsection{Spectral index analysis}
  \label{subsection.spectralindexanalysis}
    \begin{figure}[h]
      \centering
      \begin{tabular}{c}
              \includegraphics[width=0.49\textwidth]{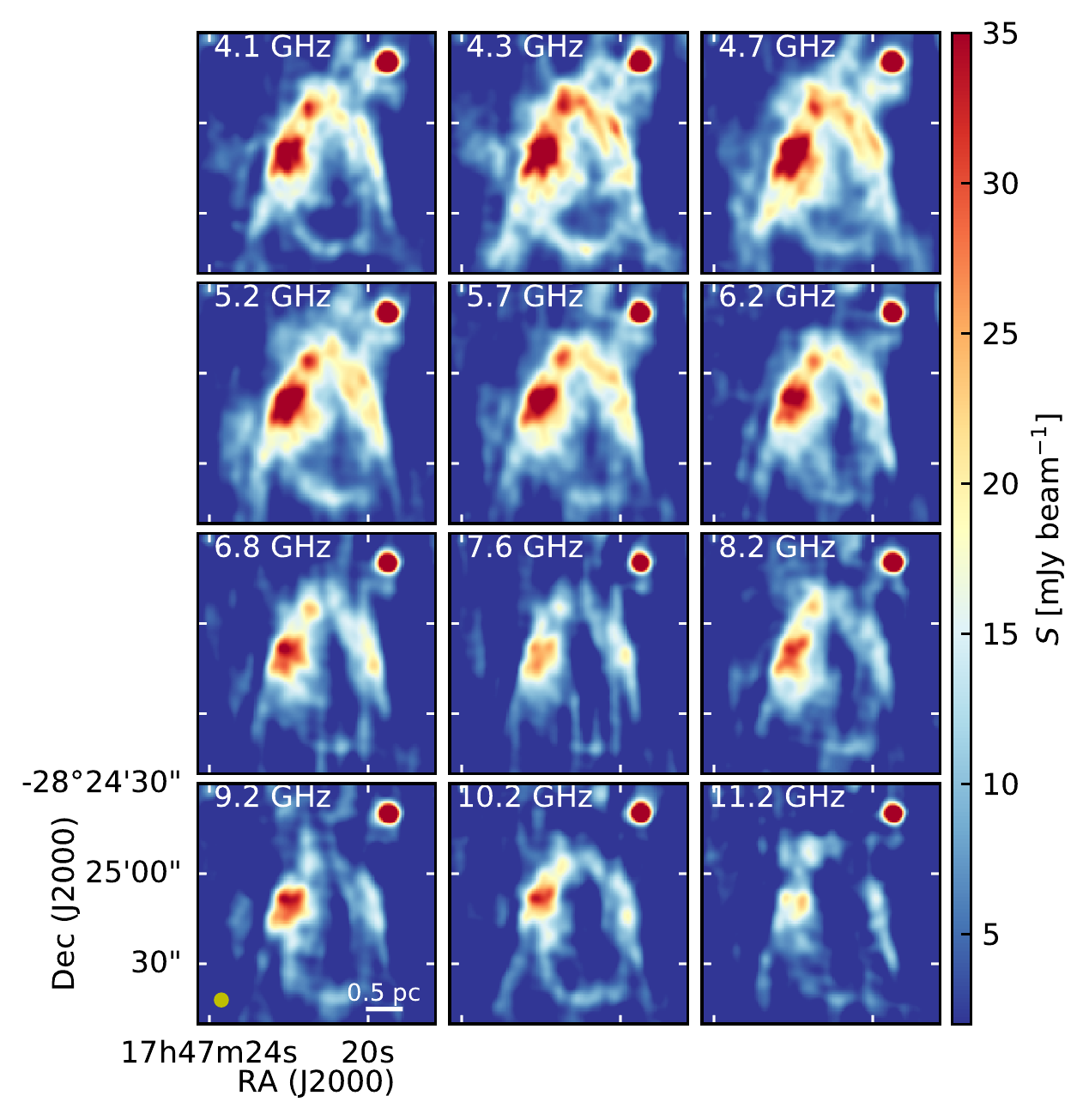}
      \end{tabular}
      \caption{Channel maps \moda{of the Sgr\,B2(DS)} region. 
      \moda{All the 12 maps have been produced considering the same $uv$ range limited to 0.6--50~k$\lambda$, and have been convolved to a circular beam of 4\arcsec. The synthesized beam is shown as a yellow circle in the bottom left panel.}}
      \label{f:chanmaps}

    \end{figure}

    \begin{figure*}[h]
      \begin{center}
      \begin{tabular}{c}
              \includegraphics[width=0.99\textwidth]{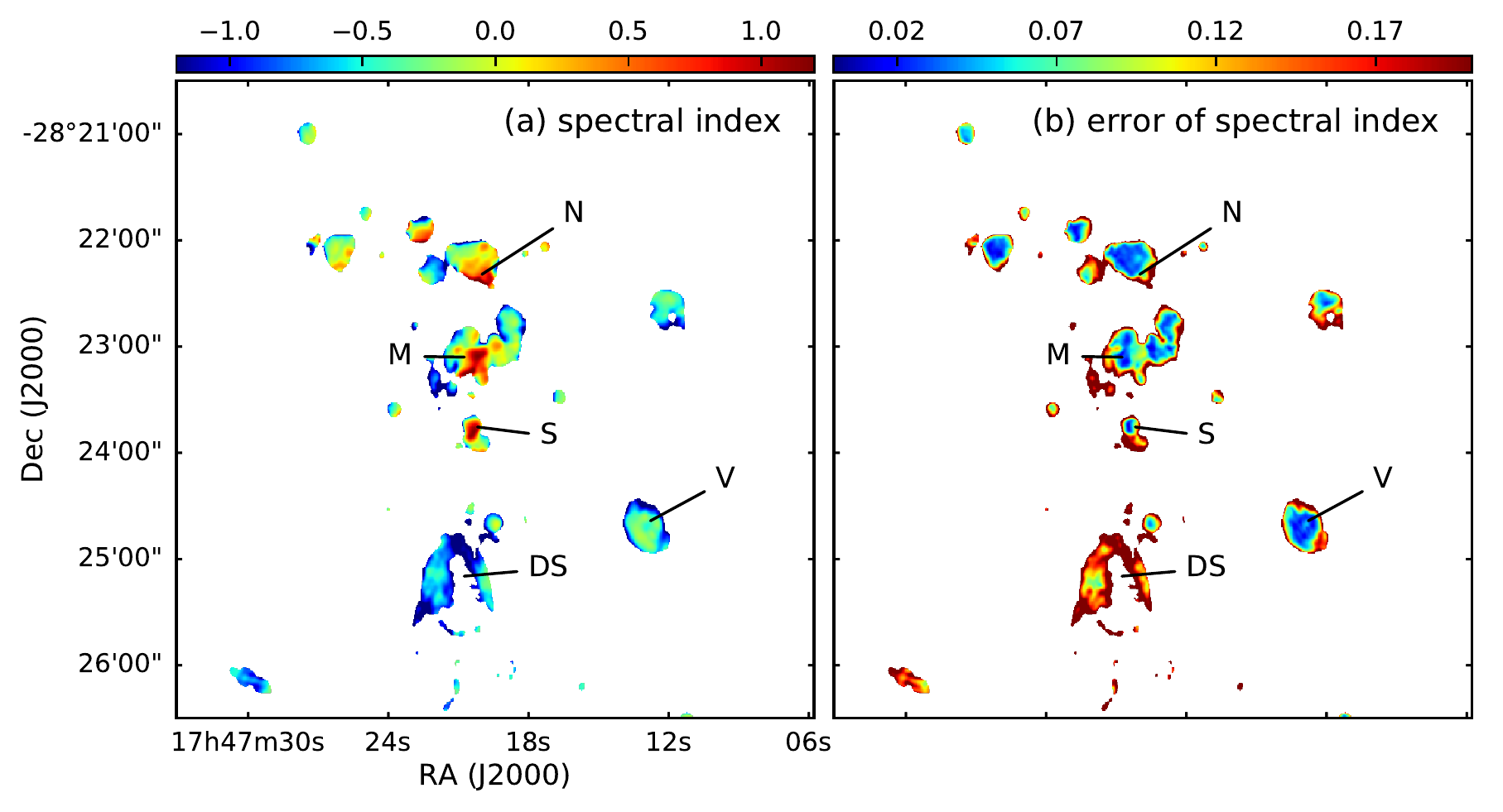} \\
      \end{tabular}
      \caption{The spectral index ($\alpha$\moda{, panel a}) and its uncertainty\moda{, (panel b) throughout}  the whole Sgr\,B2 region. \modb{Panel (a) is the $\alpha$ map of entire Sgr\,B2 region. Known \hii regions are marked. Panel (b) is the map of error of $\alpha$.}\moda{The regions marked in both panels correspond to those regions also labeled in Fig.~\ref{f:c-x-band}. See Sect.~\ref{subsection.spectralindexanalysis} for details in the calculation of the spectral index.}}
      \label{f:spix_err}
      \end{center}
    \end{figure*}

    The C and X band data, from 4 to 12~GHz, are divided into 12 tomographic maps and convolved to a common angular resolution of 4\arcsec. 
    \moda{To ensure that each of the 12 channel maps is sensitive to similar spatial scales, we have set the \textit{uv} limit to the common range 0.6--50~k$\lambda$. Moreover, we have checked that large-scale background emission does not affect the spectral analysis. For this, we have inspected intensity profiles of the 12 channel maps, and found the background emission to be zero (see Appendix~\ref{background} and Fig.~\ref{f:backgroundcheck}).}
    In Fig.~\ref{f:chanmaps}, we show the portion of the map corresponding to Sgr\,B2(DS). The intensity decreases significantly from 4~GHz to 12~GHz. We analyzed the SED of the radio emission in DS by fitting a power law to the 12 tomographic maps. We define the variation of flux with frequency as $S_\nu \propto \nu^{\alpha}$, where $S_\nu$ denotes the flux density,  $\nu$ stands for frequency, and $\alpha$ is the spectral index. The power-law fitting is conducted pixel by pixel for the whole map. For each pixel, the spectral index is fit only if the emission in all the 12 tomographic maps is above $3\sigma$, otherwise the pixel is masked. To avoid effects of possible artifacts, the fitting minimizes a loss function $  r(z) = \sqrt{1 + z/0.01} - 1 $, where $z$ is the square of the difference between the fit power-law function and the observed intensity in the 12 frequency ranges. The uncertainty of $\alpha$ is obtained from the corresponding diagonal element of the covariance matrix of the fitting. 

    \begin{figure*}[]
      \begin{center}
      \begin{tabular}{c}
              \includegraphics[width=0.99\textwidth]{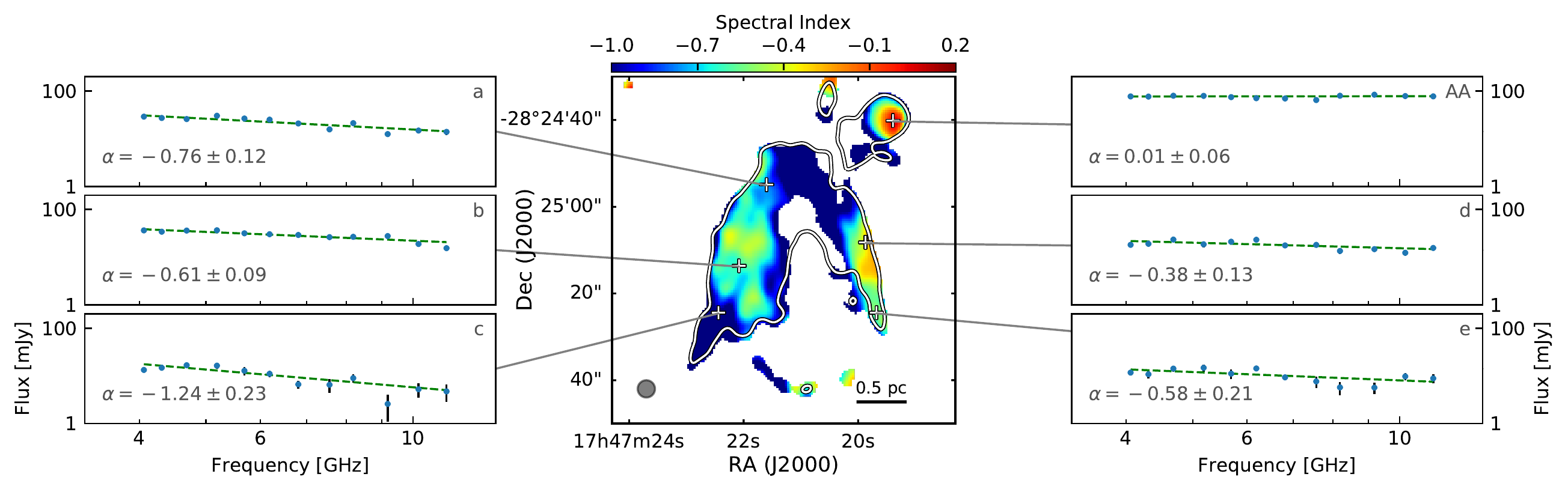} \\
      \end{tabular}
      \caption{The spectral index of DS. Six spots are taken as examples to show the fitting of SED and the fit spectral index ($\alpha$). The contours are where the flux density at 4~GHz is 10~mJy/beam. The angular resolution is 4\arcsec$\times$4\arcsec, and the beam is shown in the lower left corner.}
      \label{f:spix_ds}
      \end{center}
    \end{figure*}

    The fit spectral index map of the whole Sgr\,B2 region, as well as the map of uncertainty of $\alpha$ are shown in Fig.~\ref{f:spix_err}. The hot cores N, M, as well as other sources in the whole Sgr\,B2 region, except DS, show spectral indices between $-0.01$ and 2, which are consistent with thermal emission from H{\sc ii} regions. The central parts in N and M have spectral index values greater than 1, which indicates that the emission is optically thick. At the edges of these hot cores we determine flatter SEDs (e.g., $\alpha=-0.1$), suggesting that the centimeter emission at the edges of these H{\sc ii} regions becomes optically thin.

    In contrast to most regions in Sgr\,B2, the major part of DS shows negative spectral index values (see also Fig.~\ref{f:spix_ds}). We have highlighted the SED of six selected positions in the DS region and shown them in Fig.~\ref{f:spix_ds}. Source AA is included as reference \moda{for the spectral index}, since it is a known \hii region with optically thin free-free emission \citep{Mehringer:1993aa}\modb{, thus can be adopted as a reference for the spectral index of DS}. As displayed in Fig.~\ref{f:spix_ds}, source AA has $\alpha = 0.01\pm0.06$, consistent with optically thin free-free emission. The other five spots have negative spectral indices, ranging from $-0.4$ to $-1.2$. Such values indicate the existence of non-thermal emission in DS. As shown in the map, the distribution of the non-thermal emission \moda{extends} to more than 1~pc and forms arc structures.

    \moda{
With the goal of confirming the spectral index values derived from the VLA data, we have used data from the Giant Metrewave Radio Telescope (GMRT) at 350~MHz (Meng et al., in prep). We have recently observed a large region in the Galactic center around SgrB2, reaching an angular resolution of 12.2\arcsec$\times$11.7\arcsec (${\rm PA}=57^\circ$) and an rms noise level of 0.05 mJy/beam. In Appendix~\ref{appendix_GMRT} we compare the flux of DS at 350~MHz with the VLA data. The spectral index between 350~MHz and 4~GHz (see Fig.~\ref{f:gmrt_mixed_model}) is consistent with the VLA spectral index map shown in Fig.~\ref{f:spix_err}, confirming the presence of non-thermal emission in DS.}

  \subsection{Radio recombination line emission} 
    \label{sub:rrls}

    In addition to the continuum emission, we also observed RRLs in the whole Sgr\,B2 region.
    Due to the low sensitivity of these maps, we have smoothed the images to 8\arcsec. At this resolution we are still capable to resolve the structure of the DS region. On the smoothed maps, Gaussian fitting was conducted and peak intensity, centroid velocity, line width and integrated intensity were obtained for all the four \moda{stacked} RRLs (see Fig.~\ref{f:rrl_all}).

    The integrated intensity maps of the four RRLs show distinct spatial distribution. At 4.4~GHz, the emission appears relatively diffuse, while at 6.8~GHz, different clumpy structures are visible. At 8.9~GHz the northern part of DS is brighter, while at 10.5~GHz, the southern part of DS has more emission. Such a variation is inconsistent with a simple thermal RRL scenario but suggests the presence of other excitation mechanisms, which will be discussed in Sect.~\ref{subsec:stimulated_rrls}. The velocity maps of the four RRLs are in agreement for the four images. The eastern lobe of the bubble shows a velocity gradient ranging from velocities about 70 $\rm{km\ s^{-1}}$ in the center down to 55 $\rm{km\ s^{-1}}$ in the outer edge. As shown in the line width maps, the RRLs are typically broad, with values above $30\ \rm{km\ s^{-1}}$, significantly exceeding the thermal broadening of RRLs in H{\sc ii} regions with electron temperatures $T_{\rm e} = 10^4$~K ($\Delta v \approx 20\ \rm{km\ s^{-1}}$). Notably, the line width at the eastern edge of the bubble reaches more than $40\ \rm{km\ s^{-1}}$. The velocity difference between the eastern edge of DS and the center, together with the increase of line width at the eastern edge, suggests a possible interaction between the expanding H{\sc ii} region and its surrounding material. 


\section{Thermal and non-thermal components in Sgr\,B2(DS)} 
  \label{sec:thermal_and_non_thermal_contribution}

  The power-law fitting of the continuum emission in Sgr\,B2(DS) results in a spectral index $\alpha$ that varies from $-1.2$ to $-0.1$.  The observed range of values of $\alpha$ implies that the centimeter continuum emission in DS is a mixture of thermal and non-thermal contributions. In this section, we dissect the thermal and non-thermal components in the continuum emission and also analyze the properties of the RRL emission.

  \subsection{Disentangling the thermal and non-thermal components} 
    \label{sub:dissecting_the_thermal_and_non_thermal_components}

    In the following we present two different approaches to disentangle the contributions of the thermal and non-thermal emission in order to better characterize their origin in Sect.~\ref{sec:discussion}. The two methods are extrapolating  high-frequency emission and fitting the SED with fixed spectral indices.

    \subsubsection{Extrapolating high-frequency emission}
      \label{sec:extrapolate}

      Thermal emission at radio wavelengths is characterized by a relation in which the intensity increases with or is independent of the frequency. On the contrary, the non-thermal emission is characterized by the intensity decreasing with frequency. This suggests that the emission at higher frequencies (corresponding to 11.2~GHz in our dataset) is likely to be dominated by the thermal component, while the emission at lower frequencies (corresponding to 4~GHz) is dominated by the non-thermal component. In our first approach, we assume that the emission at the highest frequency in our data is dominated by pure thermal (free-free) emission.

\begin{figure*}[h]\
        \begin{center}
        \begin{tabular}{c}
                \includegraphics[width=0.7\textwidth]{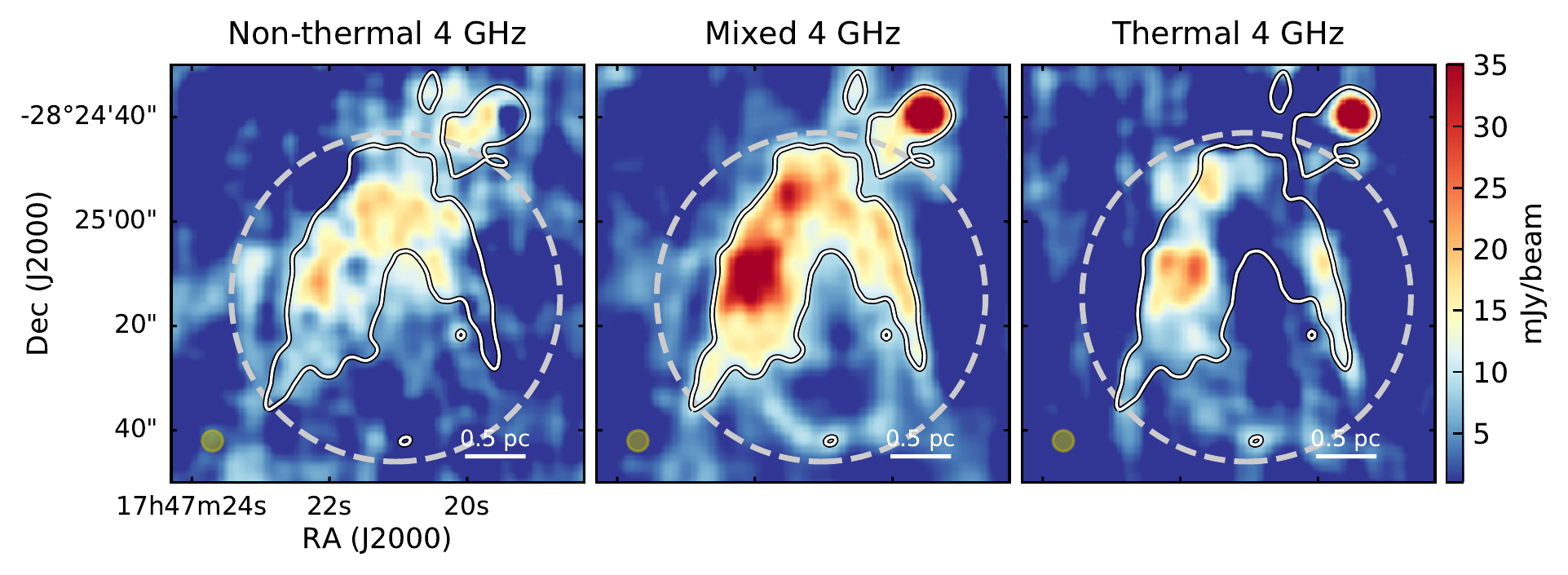} \\
        \end{tabular}
        \caption{\moda{Spatial} distribution of  synchrotron \moda{(or non-thermal, left panel)}, mixed \moda{(central panel, corresponding to observed image)} and free-free \moda{(or thermal, right panel)} component\moda{s} of DS at 4~GHz \moda{as derived from method described in Sect.~\ref{sec:extrapolate}}. \modb{The mixed emission is observed.} The thermal emission map is extrapolated from 12~GHz \moda{assuming that at this frequency the emission is originated by a thermal component with a spectral index $\alpha = -0.1$.} The non-thermal emission is obtained by subtracting the extrapolated thermal emission from the mixed emission. The contours are the same as in Fig.~\ref{f:spix_ds}.  \moda{All panels have a circular beam of 4\arcsec\ (shown in the bottom left corner of each panel). The dashed circles indicate the region in which we calculate the flux density of DS (see Sect.~\ref{sub:dissecting_the_thermal_and_non_thermal_components}).}}
        \label{f:spixtomo}
        \end{center}{}
      \end{figure*}

      The total flux of DS at 11.2~GHz is 0.5~Jy within a diameter of $\sim$36\arcsec, corresponding to a brightness temperature of 14~K. For a typical \hii region temperature of $5\times10^3$--$10^4$~K, the optical depth $\tau$ ranges from $1.4\times10^{-3}$ to $2.8\times10^{-3}$. Therefore, the free-free emission of DS is optically thin. We use the typical spectral index $\alpha = -0.1$ of optically thin free-free emission to extrapolate the 11.2~GHz flux density to 4~GHz. The extrapolated thermal component is subtracted from the observed flux density at 4~GHz to get a pure non-thermal component. In Fig.~\ref{f:spixtomo} we show the derived thermal and non-thermal components at 4 GHz. The non-thermal emission appears more widespread, while the thermal component appears concentrated in different clumps located along the edge of the bubble.  As expected, the \hii region AA has strong thermal emission, while at 4~GHz, the contribution from the non-thermal component drops below the RMS level. Only in the southeastern region, connecting source AA with the bubble-like structure of DS, we find some presence of a possible contribution of non-thermal emission.

      \moda{The total observed flux at 4 GHz (within the circle highlighted in Fig.~\ref{f:spixtomo}) is $\sim$1.5~Jy, for which we determine that $\sim$60\% has a non-thermal origin. We note that since we have assumed that the emission at 11.2~GHz is purely thermal, the contribution of thermal emission at 4~GHz is most likely overestimated and therefore, the non-thermal contribution may be underestimated.}

    \subsubsection{Fitting the SED with fixed spectral indices}
    \label{subsub:fittingwithtwoseds}

      In a second approach, we aim at simultaneously determining  the contribution of  thermal and non-thermal emission. For this, we fit the SED covering the whole range from 4 to 12 GHz with two power-law functions describing each component. For the thermal component, the power-law is $S_{\rm{th}}(\nu) \propto \nu^{-0.1}$, assuming that the emission is optically thin free-free. For the non-thermal component, we use $S_{\rm{nt}}(\nu) \propto \nu^{-0.7}$ \citep[see][]{Hollis:2003aa,Protheroe:2008aa,Jones:2011aa}. We fit the observed SED with a linear superposition of these two power-law functions and get the contribution of each component at different frequencies.

      \begin{figure*}
        \begin{center}
        \begin{tabular}{c}
                \includegraphics[width=0.7\textwidth]{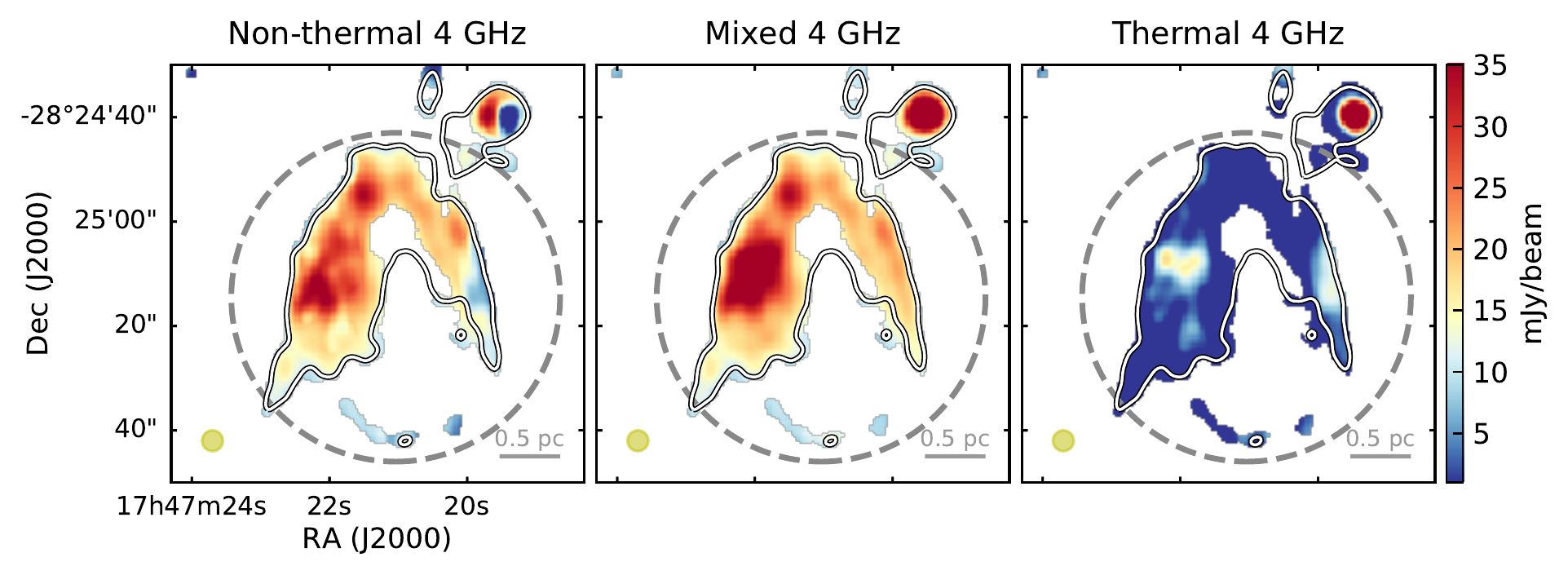} 
        \end{tabular}
        \caption{\moda{Spatial} distribution of the synchrotron \moda{(left panel)}, mixed \moda{(central panel)} and free-free \moda{(right panel)} component\moda{s} of DS at \moda{4~GHz}. \modb{The mixed emission is the observed continuum.} \moda{All the three images are obtained after fitting the observed SED with the sum of two power-law functions:} $S_{\rm{th}}(\nu) \propto \nu^{-0.1}$ and $S_{\rm{nt}}(\nu) \propto \nu^{-0.7}$ for \moda{the} thermal and non-thermal components\moda{,} respectively (see details in Sect.~\ref{subsub:fittingwithtwoseds}). The contours are the same as in Fig.~\ref{f:spix_ds}. \moda{All panels have a circular beam of 4\arcsec\ (shown in the bottom left corner of each panel).} The dashed circles \moda{indicate} the region in which we calculated \moda{the} flux density of DS \moda{(see Sect.~\ref{sub:dissecting_the_thermal_and_non_thermal_components})}.}
        \label{f:spixtomo2}
        \end{center}
      \end{figure*}

      As shown in Fig.~\ref{f:spixtomo2}, at 4~GHz, the spatial distribution of the non-thermal component is more widespread compared to the thermal component \moda{which} is mainly concentrated in the central part of the eastern lobe\moda{. This is }consistent with the spatial distribution of the two components shown in Sect.~\ref{sec:extrapolate}. The \hii region AA appears to have pure thermal emission in the center, with a possible contribution of non-thermal emission in the outskirts. The presence of non-thermal emission in the outskirts of \hii regions have been found in a handful of objects \citep[e.g.,][]{Garay:1996aa,Mucke:2002aa,Veena:2016aa,Veena:2019aa}.

      \moda{
      From the simultaneous fit of a thermal and non-thermal components, we determine that 90\% of the total emission at 4~GHz has a non-thermal origin in Sgr\,B2(DS). 
      The difference of the relative contribution of thermal and non-thermal components derived from the two methods (see Sect.~\ref{sec:extrapolate}) can be due to either the assumption of the thermal dominance at 11.2~GHz considered in the model presented in the previous section,  or the fixed values of $\alpha$ for the thermal and non-thermal components used in this method. Overall, the results of both methods are in agreement and confirm the presence of extended non-thermal emission in Sgr\,B2(DS). Further observations at higher and lower frequencies may help to better constrain the properties and distribution of the thermal and non-thermal components in this region.}


  \subsection{Stimulated RRLs}
  \label{subsec:stimulated_rrls}

\begin{figure*}[h]
      \begin{center}
      \begin{tabular}{c}
              \includegraphics[width=0.7\textwidth]{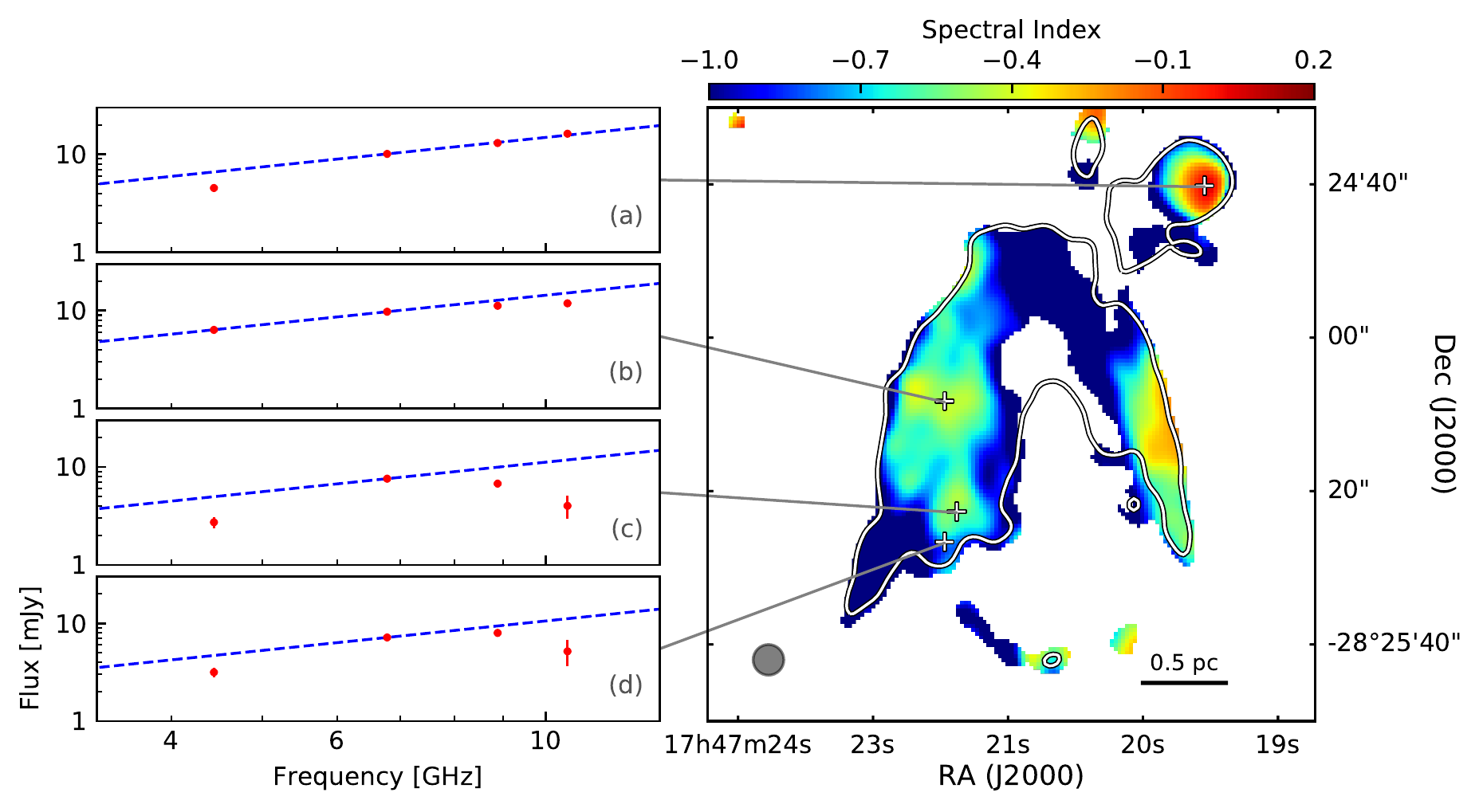}
      \end{tabular}
      \caption{\textit{Right:} \moda{Spectral index map of the Sgr\,B2(DS) as shown in the central panel} of Fig.~\ref{f:spix_ds}. \modc{The map of the uncertainty of spectral index is as shown in the right panel of Fig.~\ref{f:spix_err}.} \textit{Left:} \moda{Peak intensities of the four stacked RRLs (see Table~\ref{t:rrls}) at four selected positions. Position (a) corresponds to the well-known H{\sc ii} region AA, while the other positions have been selected to probe regions with negative spectral indices. The dashed line in each panels marks the scenario for LTE under which the flux is proportional to the frequency.}}
      \label{f:rrl_stim}{}
      \end{center}
    \end{figure*}

    As discussed in Sect.~\ref{sub:rrls}, the RRLs integrated intensity distribution varies significantly among the four stacked frequencies. From Fig.~\ref{f:rrl_all}, we can see that the peak intensity of RRLs at the center of the eastern lobe increases monotonically with frequency, which suggests that the RRLs at this part of DS are excited under local thermodynamic equilibrium (LTE) conditions, since under LTE the peak intensity of RRLs obey $S_{\nu} \propto \nu$. However, at the edge of DS, as shown in Fig.~\ref{f:rrl_stim}, the peak intensities of the RRLs exhibit anti-correlation with frequency at 6.8, 8.9, and 10.5~GHz, which suggests that the RRLs in these regions are probably under non-LTE conditions.

    \begin{figure*}[h]
      \begin{center}
      \begin{tabular}{c}
              \includegraphics[width=0.9\textwidth]{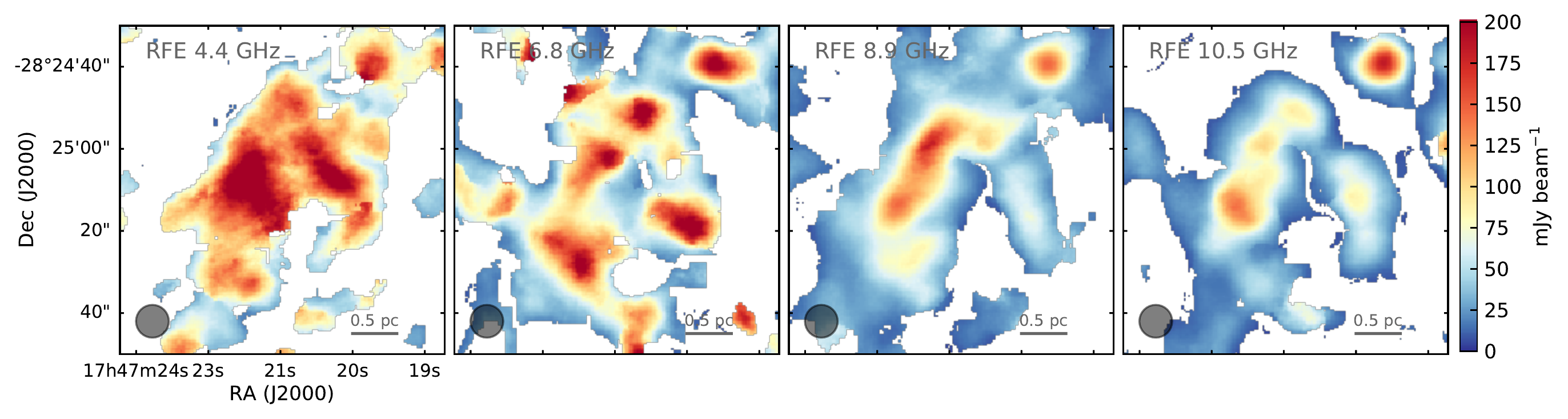}\\
              \includegraphics[width=0.9\textwidth]{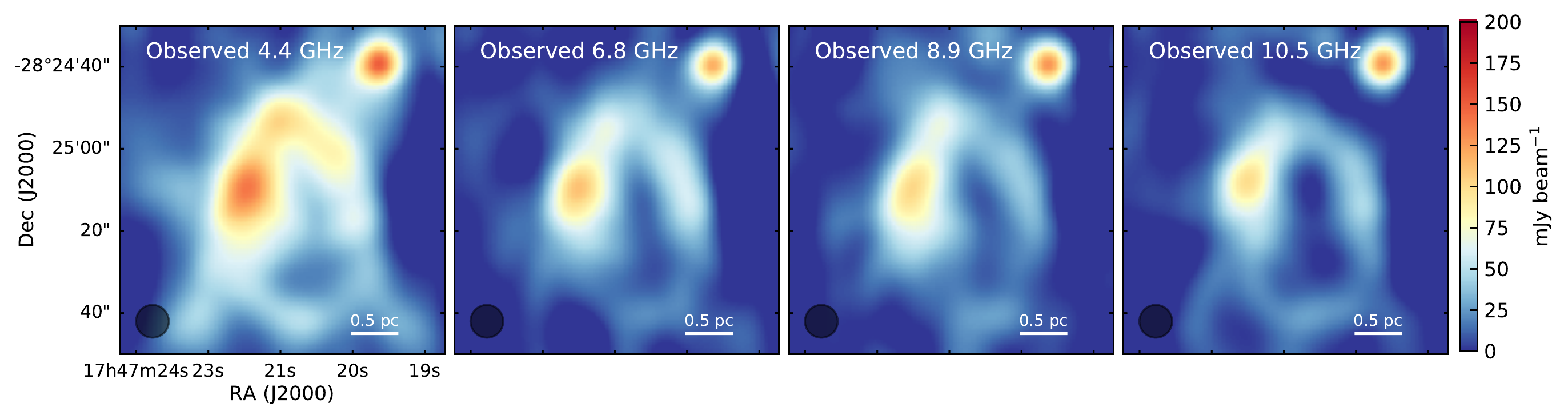}
      \end{tabular}
      \caption{\textit{\moda{Top panels}}: \moda{Free-free} continuum maps of the Sgr\,B2(DS) region derived from the four staked RRLs (RFE \moda{as described in Sect.~\ref{subsec:stimulated_rrls}, following Eq.~\ref{EQ:T_L/T_C}}) at 4.4, 6.8, 8.9, and 10.5~GHz.\modb{, by applying equation (\ref{EQ:T_L/T_C})}. \textit{\moda{Bottom panel}}: \moda{Observed} continuum emission at 4.4, 6.8, 8.9 and 10.5~GHz. For comparison, the observed continuum maps are also smoothed to 8\arcsec, the beam is shown as the shaded circle in the lower left of each plot.}
      \label{f:4rrls_cont}
      \end{center}
    \end{figure*}

    We also find that the RRLs are likely excited under non-LTE conditions when comparing their emission with the continuum brightness. Under LTE, the integrated intensity of RRLs relates to optically thin free-free emission as
    \begin{equation}
          \frac{S_{\rm L}}{S_{\rm C}}\left(\frac{\Delta v}{\rm km\ s^{-1}}\right)
          =
          6.985\times10^3 \left(\frac{\nu}{\rm GHz}\right)^{1.1}
          \left(\frac{T_{\rm e}}{\rm K}\right)^{-1.15},
          \label{EQ:T_L/T_C}
    \end{equation}
    where $S_{\rm L}$ and $S_{\rm C}$ are the peak flux of RRL and flux of free-free continuum, respectively, and $T_{\rm e}$ is the electron temperature. Adopting $T_{\rm e} = 8000\ \rm{K}$ \citep[see ][]{Mehringer:1993aa}, we derive the corresponding free-free continuum level of the four stacked RRLs (RRL-derived free-free emission, hereafter RFE). The comparison between the RFE and the continuum emission at the four frequencies is shown in Fig.~\ref{f:4rrls_cont}. The observed continuum emission is supposed to be a mixture of free-free and synchrotron emission, which should have a higher intensity than the RFE maps. However, the RFE exceeds the observed continuum by a factor of $\sim$2 at the low frequency end. One possible origin of this excess is that the RRLs are not under LTE conditions but stimulated.

\begin{figure}[h]
      \begin{center}
      \begin{tabular}{c}
              \includegraphics[width=0.48\textwidth]{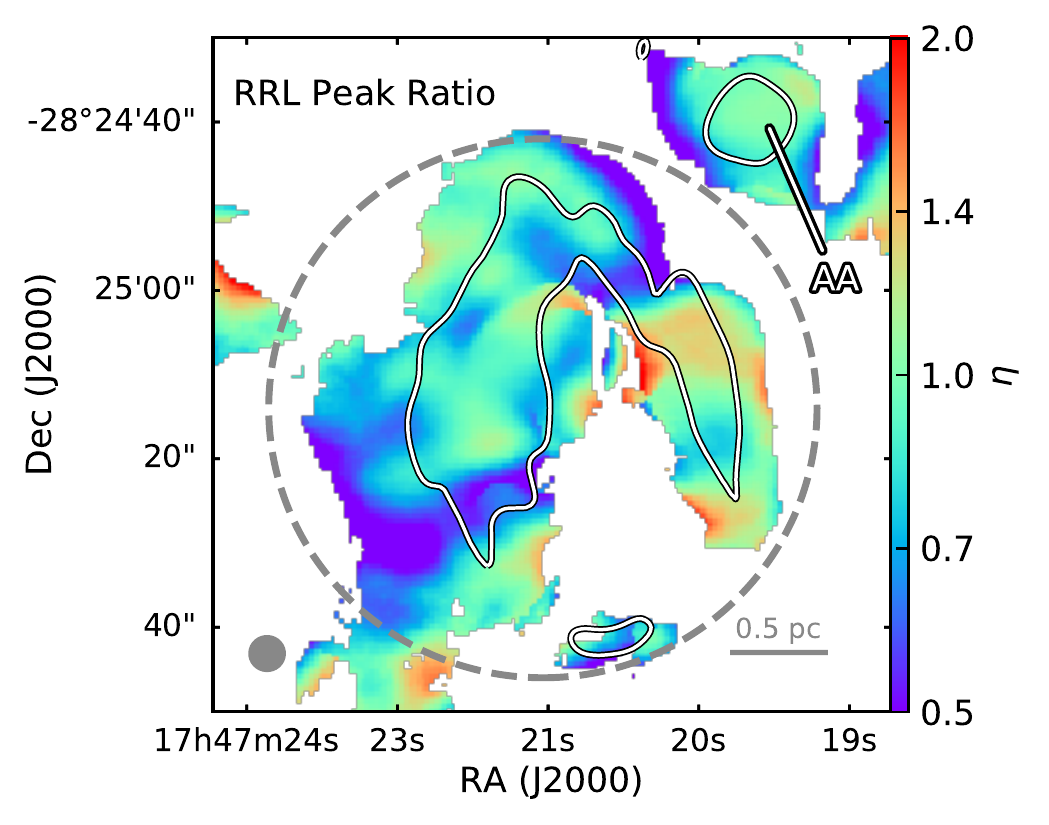}
      \end{tabular}
      \caption{ Map of $\eta$ \moda{(see Sect.~\ref{subsec:stimulated_rrls})} in DS. The X band continuum emission is overlaid as contours. \moda{The synthesized beam, corresponding to 8\arcsec,} is plotted as \moda{a} dark circle at the lower left \moda{corner}. The dashed circle indicates the area in which pixels are taken into account for Fig.~\ref{f:etaalpha}.}
      \label{f:eta}{}
      \end{center}
    \end{figure}

    To characterize the stimulation of RRLs, we introduce the RRL peak ratio $\eta$. 
    According to theoretical models, the peak intensity of stimulated RRLs and frequency are anti-correlated at high frequencies \citep{Shaver:1978ab}. Therefore, we consider the RRLs at 8.9 and 10.5~GHz and define
    \begin{equation}
      \eta = 0.85 \frac{S_{10.5~\rm GHz}}{S_{8.9~\rm GHz}},\label{eq:eta}
    \end{equation} where $S_{10.5~\rm GHz}$ and $S_{8.9~\rm GHz}$ are the RRL peak intensities, respectively, while the normalization factor $0.85 = {8.9}/{10.5}$ result in $\eta = 1$ when LTE conditions hold. When the RRLs are stimulated, $\eta <1$. In Fig.~\ref{f:eta} we show the $\eta$ map of DS.  From the plot, we see that in the central and western part of DS, $\eta\approx1$, which indicates that the emission is under LTE. At the eastern edge of DS, we find $\eta<1$, which suggests the presence of stimulated emission. Notably, source AA displays $\eta = 1$, meaning that its emission is under LTE, which is consistent with the scenario that source AA is an H{\sc ii} region with only thermal emission.

    \begin{figure}[h]
      \begin{center}
      \begin{tabular}{c}
              \includegraphics[width=0.48\textwidth]{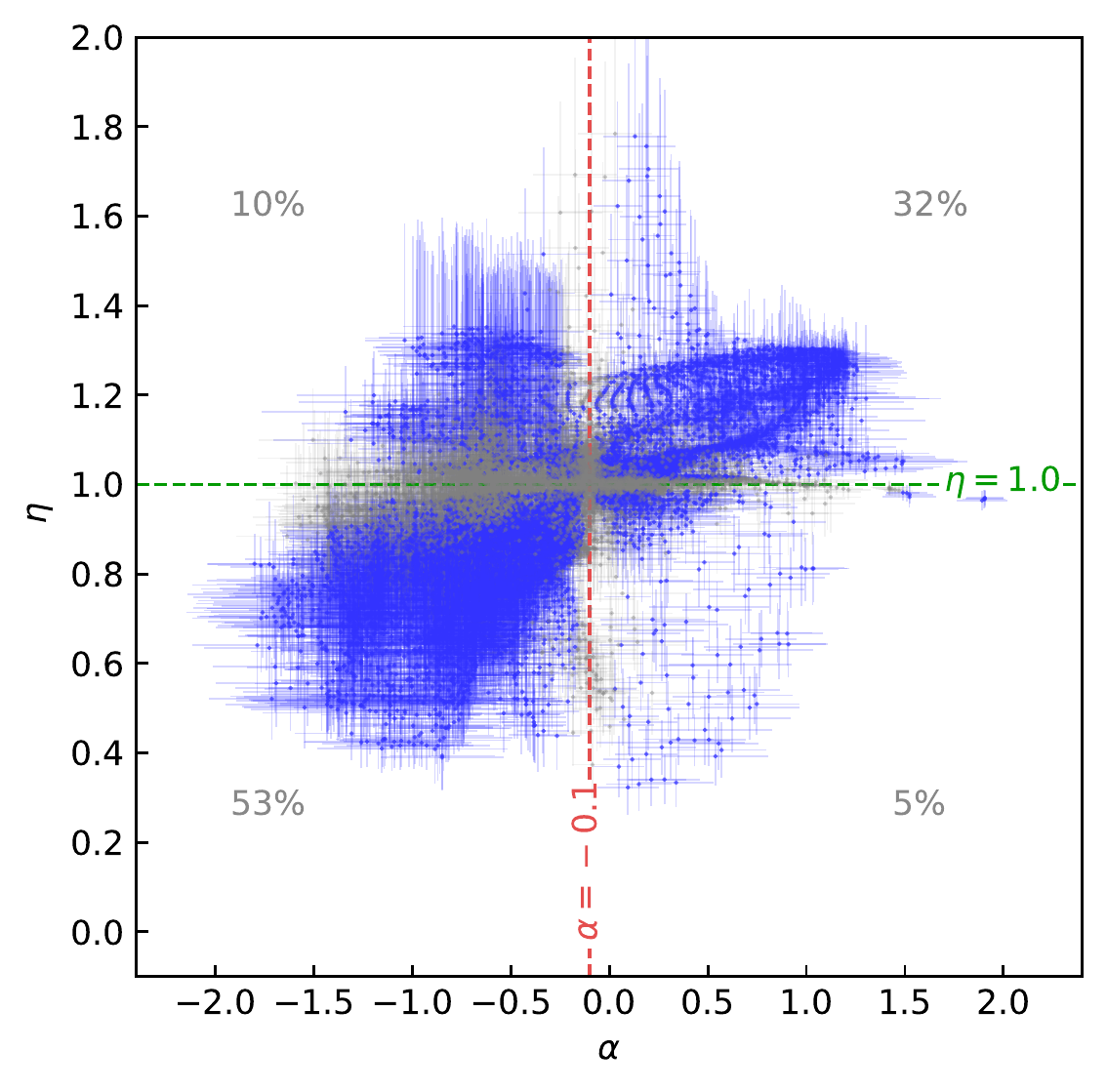}
      \end{tabular}
      \caption{\moda{Plot} of $\eta$  against spectral index ($\alpha$). Only the pixels that are in the area marked by the dashed circle in Fig.~\ref{f:eta} are taken into account, thus the influence of source AA is excluded. The criterion of non-thermal emission, $\alpha=-0.1$, is shown \moda{with a} vertical dashed line, while the criterion of stimulated \moda{RRL emission}, $\eta=1$\moda{,} is shown as \moda{a} horizontal dashed line. These two orthogonal lines divide the whole $\alpha$--$\eta$ space into four quadrants. For each quadrant, the number of pixels within it as a percentage of the number of all the pixels is written. The points with error bars crossing the $\alpha=-0.1$ or $\eta=1$ lines are colored gray and neglected in the statistics.
      }
      \label{f:etaalpha}{}
      \end{center}
    \end{figure}

    We quantify the correlation between non-thermal emission and the stimulation of RRLs, by plotting $\eta$ against $\alpha$ pixel by pixel (see Fig.~\ref{f:etaalpha}).  A total of 63\% of the pixels in DS have $\alpha<-0.1$, meaning associated with non-thermal emission. Among these non-thermal pixels, 86\% (53\% of all the pixels in DS) have $\eta<1$ , i.e., the RRLs are stimulated under non-LTE conditions. Out of all the pixels, 58\% have $\eta<1$. The vast majority (92\%) are associated with non-thermal emission. Therefore, most of the area in DS associated with non-thermal emission is also associated with stimulated RRLs, and similarly, almost all the parts associated with stimulated RRLs show non-thermal emission. The correlation between $\alpha$ and $\eta$ seen in Fig.~\ref{f:etaalpha} is consistent with the scenario that RRLs can be stimulated by non-thermal emission \citep{Shaver:1978ab}. 

\section{Origin of thermal and non-thermal emission} 
  \label{sec:discussion}

  We have found that the DS region, located within the envelope of Sgr\,B2, shows a mix of thermal free-free and non-thermal synchrotron continuum emission at radio wavelengths. This kind of radio continuum emission may have different origins: \hii regions, supernova remnants (SNR\moda{s}), planetary nebulae (PNe)\modc{ or extragalactic radio sources.} Although the most probable origin is the \hii region, we have explored the other \modc{three} scenarios. We searched the XMM-Newton catalog for X-ray sources in the region, and found no source associated with DS in the 0.5--12~keV continuum bands \citep{Ponti:2015aa}. The lack of bright X-ray emission suggests that we can exclude \moda{a} scenario in which DS is a SNR. In a different scenario, PNe can emit in the radio regime, however, their typical size ranges from 0.03 to 0.1~pc, i.e., significantly smaller than the radius of DS (0.5~pc). Additionally, the age of PNe is expected to be between $10^{7}$ and $10^{10}$~yr \citep[e.g.,][]{Bressan:1993aa}, much longer than the age of Sgr\,B2  \citep[estimated to be about 0.74~Myr, see][]{Kruijssen:2015aa}. Also, we estimate the total mass of the ionized gas in DS to be $\sim 500\ M_{\odot}$, which is much larger than the mass of a typical low-mass star that can generate a PN. Therefore, we exclude the possibility that DS is a PN. \modc{The minimum detectable flux density of our images is $\sim$0.2~mJy, which can be used to estimate the expectable number of  extragalactic radio sources. In the entire area of Sgr\,B2 (7\arcmin$\times$7\arcmin), the expected number of extragalactic radio sources is $\sim$1, while in the 1\arcmin$\times$1\arcmin area of DS, the number is $\sim$0.03 \citep[see][]{Condon:1998aa,Anglada:1998aa}. Additionally, the spatial extension of DS ($\sim$1\arcmin) is significantly larger than the typical size of extragalactic radio sources \citep[see][]{Condon:1998aa}. Therefore, the possibility that DS is an extragalactic radio source is excluded. Thus,} the only remaining scenario is DS being an \hii region. In this section, we discuss the properties of the central star and the possible mechanisms that can produce the observed non-thermal emission. 
{}

  \subsection{Ionization by a central star} 
    \label{sub:thermal_mechanism}

    The flux of Lyman continuum photons, $\dot{N}_\mathrm{Ly}$, needed to ionized an \hii region can be derived as following \citep{Schmiedeke:2016aa}:
    \begin{equation}\label{eq:lyman_photons}
      \frac{\dot{N}_\mathrm{Ly}}{\rm{s^{-1}}} = 8.9\times10^{40} \frac{S_{\nu}}{\rm{Jy}} \left( \frac{\nu}{\rm{GHz}} \right)^{0.1} \left( \frac{T_{\rm{e}}}{\rm{10^4\ K}} \right)^{-0.45} \left(\frac{D}{\rm{pc}}\right)^2,
    \end{equation} 
    where $S_{\nu}$ is the flux at frequency $\nu$, $T_{\rm{e}}$ is the electron temperature, and $D$ is the distance to the source. The total flux of the free-free emission in DS is 0.5~Jy at 11.2~GHz, assuming that the continuum emission at this frequency is pure thermal. For a typical \hii region, the electron temperature ranges from 5000 to 10000~K \citep{Zuckerman:1967aa}. Taking $D = 8.34 \ \rm{kpc}$ for Sgr\,B2, we obtain a flux of Lyman continuum photons of 4--6~$\times10^{48}\ \rm{s^{-1}}$. Such a Lyman continuum flux corresponds to an O7 ZAMS star \citep[see Table~II of ][]{Panagia:1973aa}. We have searched the \textit{Spitzer} infrared images \citep[e.g.,][]{Ramirez:2008aa} for a possible infrared counterpart of the star ionizing the \hii region in DS, but found no clear candidate. \modb{likely due to the high extinction in the infrared bands towards the Galactic center.} \moda{
    Also the young stellar object catalog of \citep{Yusef-Zadeh:2009aa} does not show any infrared source in the center of DS. One possible reason for the lack of a detected infrared source in DS is the high extinction. Assuming a column density of molecular hydrogen of $10^{24}$~$\rm{cm^{-2}}$ in DS \citep{Schmiedeke:2016aa,Ginsburg:2018aa}, the extinction at 5~$\mu$m is $\sim 20$, based on the grain model by \citet{Li:2001aa}. For an O7 star at a distance of 8.34~kpc, this extinction result in an apparent magnitude $\sim 30$, which is non-detectable in current infrared images.}

    \begin{figure*}[h]
      \begin{center}
      \begin{tabular}{c}
              \includegraphics[width=0.95\textwidth]{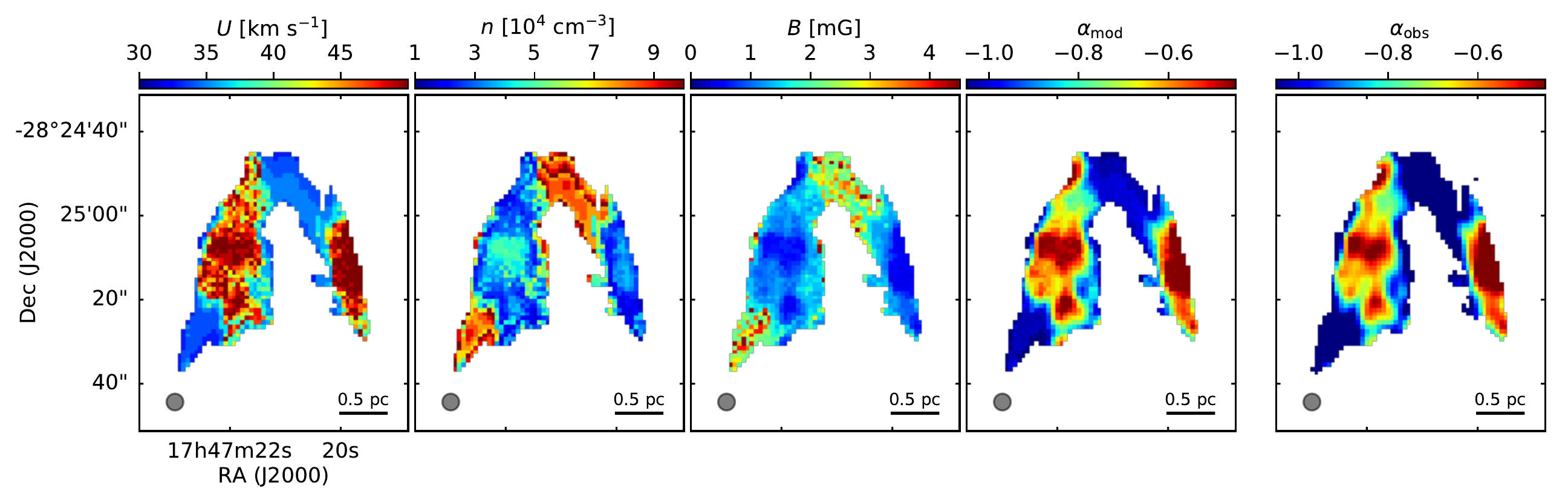}
      \end{tabular}
      \caption{\meng{\moda{Maps} of shock velocity ($U$), volume density ($n$), and magnetic field strength ($B$) of DS that reproduce the observed flux density maps at the 12 frequencies (see Fig.~\ref{f:chanmaps}) by the First-order Fermi acceleration model through a $\chi^2$ test using the model described in \citet{Padovani:2019aa}. The model also \moda{generates} the modeled spectral index ($\alpha_{\rm mod}$) map which is consistent with the observed $\alpha_{\rm obs}$ map. 
      }}
      \label{f:marcomodel}
      \end{center}
    \end{figure*}

  \subsection{Non-thermal emission origin} 
    \label{sub:non-thermal_mechanism}

    The interaction between cosmic-ray nuclei and the interstellar medium produces secondary electrons (and positrons). These secondary electrons can be a source of synchrotron emission like the one detected in Sgr\,B2(DS). \citet{Protheroe:2008aa} analyze diffusion models and indicate that for secondary electrons to radiate synchrotron emission at GHz frequenc\moda{ies} in $\sim$1~mG magnetic field, cosmic-ray particles should have multi-GeV energies, whereas such cosmic-ray particles cannot penetrate into the cloud due to the high gas density in Sgr\,B2 \citep[see, e.g.,][]{Padovani:2009aa,Padovani:2013aa}. Thus, the penetration of cosmic \moda{rays} from outside (e.g., from the nearby Galactic center) \moda{\modb{can be}} is excluded \moda{as a possible origin to the non-thermal emission observed in Sgr\,B2(DS).}

    However, the presence of synchrotron emission is the fingerprint of relativistic electrons. \citet{Padovani:2015aa,Padovani:2016aa} showed that particle acceleration can take place at a shock location in protostars according to the first-order Fermi acceleration mechanism (also known as diffusive shock acceleration). In fact, local thermal particles are accelerated up to relativistic energies while crossing back and forth a shock surface thanks to magnetic fluctuations around the shock itself.

    In a companion paper \citep{Padovani:2019aa}, we present an extension to this theory applied to \hii regions. We demonstrate that electrons can be \moda{efficiently} accelerated under the hypothesis that the \moda{flow} velocity \moda{in the reference frame} of an expanding \hii region hitting denser material is sufficiently high ($>35$~km~s$^{-1}$) to switch on particle acceleration. The main parameters of the model are the magnetic field strength, the \modb{shock velocity} \moda{the flow velocity in the shock reference frame (hereafter velocity)}, the temperature, the volume density, and the ionization fraction.  Assuming a complete ionized medium and a temperature of 8000~K (see Sect.~\ref{subsec:stimulated_rrls}), we studied the parameter space volume density-magnetic field strength and computed the emerging shock-accelerated electron flux for a shock radius of 0.36~pc, which is the average distance between the center of DS and the inner surface of the synchrotron emitting region. This electron flux is used to compute the synchrotron emissivity and then the flux density between 4 and 12~GHz for a beam of $4\arcsec$ and an average emitting region size\footnote{The average size, $L$, of a shell-shaped emitting region with inner and outer radii, $R_{\rm in}$ and $R_{\rm out}$, respectively, is $L=(\pi/2)R_{\rm out}(1-R_{\rm in}^2/R_{\rm out}^2)$. For DS we assume $R_{\rm in}=0.36$~pc and $R_{\rm out}=0.72$~pc.} of 0.85~pc. We performed a $\chi^2$ test to find the values of the velocity ($U$), density ($n$), and magnetic field strength ($B$) that best reproduce the observations. For the flux densities we obtain an average accuracy of about 20\% assuming $33\lesssim U\lesssim50$~km~s$^{-1}$, $10^4\lesssim n/{\rm cm^{-3}}\lesssim 9\times10^4$, and $0.3\lesssim B/{\rm mG}\lesssim 4$. From the model, we derive maps of the velocity, magnetic field strength and density in the Sgr\,B2(DS) region  (see Fig.~\ref{f:marcomodel}). 
    The magnetic field strength is found to be in the range 0.5--2~mG as reported by \citet{Crutcher:1996aa}. The model also gives an $\alpha$ map of DS, which is in agreement with the observed $\alpha$ map (Fig.~\ref{f:spix_ds}). It is interesting to notice that the velocity is lower where the density is higher. One reason might be that the flow is moving faster towards the east-west direction, which is the direction of minimum resistance. In contrast, the velocity is lower toward the northern direction where the density is higher.  
    \moda{Although the RRL line width cannot be directly used to determine the shock speed, we note that it is significantly larger than the typical expansion velocity of an H{\sc ii} region \citep[10~${\rm km\  s^{-1}}$][]{Draine:2011aa}, which points to an additional mechanism of acceleration.   } One possible origin of the high velocity is the presence of a stellar wind driven by the O7 star that likely ionizes the \hii region \citep[see e.g.,][]{Veena:2016aa,Pereira:2016aa,Kiminki:2017aa}. We note that velocities of about 35~km~s$^{-1}$ are in agreement with what is obtained by simulations of \hii regions of O and B stars driving strong stellar winds \citep[][]{Steggles:2017aa}.

    It is important to remark that the flux densities (from which the spectral index is derived) estimated by our model are based on the assumption that all the parameters (magnetic field strength, velocity, temperature, density, and ionization fraction) are constant along the line of sight. For a proper modeling, one should account for the spatial variations of these quantities. However, the fact that model results are already fairly comparable with observations is an indication that shock-accelerated electrons may represent the key to explain the synchrotron emission in Sgr\,B2(DS). \moda{Previous studies \citep[e.g.,][]{Yusef-Zadeh:2007aa,Yusef-Zadeh:2007ab,Yusef-Zadeh:2013aa,Yusef-Zadeh:2016aa} indicate that cosmic-ray induced non-thermal emission is common in the Galactic center region. On large scales, where the density is not as high as in the envelope of SgrB2, cosmic rays may play a dominant role in the production of relativistic electrons, while in extremely dense regions (and smaller scales, like Sgr\,B2(DS)) H{\sc ii} region shocks may be a dominant source of non-thermal emission.}


\section{Summary} 
  \label{sec:summary}{}

  We \moda{have} observed the giant molecular cloud Sgr\,B2 with \moda{the} VLA in \moda{its} D and CnB configurations. We obtained continuum and RRL maps in the frequency range 4--12~GHz, covering a spatial extent of about $20^\prime\times20^\prime$, with a beam size of 4\arcsec. We have focused our \modb{detailed} study on the Sgr\,B2(DS) region, located in the southern area of the Sgr\,B2 envelope. Our main results are:

  \begin{itemize}
    \item Sgr\,B2(DS) is bright at radio wavelengths, with intensities in the range 10--50~mJy/beam, within the observed 4 to 12~GHz frequency range. At 4\arcsec\ resolution, DS \moda{appears as} bubble-like \hii region with a diameter of about 1.5~pc, powered by an O7 star, and surrounded by dense gas and a series of dense cores distributed in an arc structure around it.

    \item We find that the total flux of DS decreases from 1.9~Jy at 4~GHz down to 0.5~Jy at 12~GHz. Spectral analysis shows that the spectral index of DS varies from $-0.4$ to $-1.2$, suggesting the presence of non-thermal emission, in addition to the thermal emission of the \hii region. We decompose the thermal and non-thermal components in DS, and find that the thermal emission is clumpy and concentrated in the eastern lobe of the bubble-like structure, while the non-thermal emission appears widespread over the whole region.

    \item \moda{The emission of RRLs in DS show} a central velocity varying from 70~$\rm{km\ s^{-1}}$ in the center to 55 $\rm{km\ s^{-1}}$ in the outer edge. The line widths of the RRLs range from  30 to 40~$\rm{km\ s^{-1}}$. From the RRLs integrated intensity, we derive the corresponding thermal continuum emission, which exceeds the observed continuum by a factor of about 2 at the low frequency end. In addition, the RRLs intensity does not follow the $S_{\nu}\propto\nu$ relation expected \moda{for} RRLs under LTE conditions, but drops at high frequencies likely due to non-LTE \moda{effects}. We find a correlation between the presence of non-thermal emission (i.e., negative spectral indices) and those regions where RRLs are excited under non-LTE conditions. This suggests that RRLs in Sgr\,B2(DS) are possibly stimulated by synchrotron emission.

    \item We modeled the observed synchrotron emission, and found that relativistic electrons can be produced via first-order Fermi acceleration, which is triggered by the interaction between the expanding \hii region and the denser surrounding material. The model, which is presented \moda{in a} companion paper of \citet{Padovani:2019aa}, reproduces the observed flux density and spectral index. From the model, we derive maps of the \moda{flow} velocity \moda{in the shock reference frame}, magnetic field strength and density in the Sgr\,B2(DS) region. 
    Velocities are found to be between 35 and \moda{50}~km~s$^{-1}$ as found \modb{in the observations of RRLs} 
    \moda{in simulations of cometary H{\sc ii} regions of O and B stars driving strong stellar winds.} The magnetic field strength is found to be in the range 0.3--4~mG as reported by \citet{Crutcher:1996aa} and the density in the range of $1-9\times10^4$~cm$^{-3}$. 

  \end{itemize}

  \moda{In this work we have characterized the non-thermal emission observed towards the H{\sc ii} region Sgr\,B2(DS), and we have proposed that locally accelerated particles, under certain conditions of density, magnetic field, and shock velocity, can be the origin of the non-thermal emission. This result raises the question whether other H{\sc ii} regions may also show non-thermal emission. A handful of H{\sc ii} regions have been reported to have negative spectral indices suggestive of a non-thermal component \citep[e.g.,][]{Garay:1996aa,Veena:2016aa,Veena:2019aa}, indicating that the presence of non-thermal emission in H{\sc ii} regions may be a more common event than previously thought. The advent of extremely sensitive facilities at radio wavelengths such as SKA (Square Kilometer Array) or the ngVLA (next generation Very Large Array) may trigger the systematic search for H{\sc ii} regions with non-thermal emission, thus allowing a study of the conditions under which non-thermal emission exists.}

\begin{acknowledgements}
  \moda{We thank the referee for his/her comments during the reviewing process.}. FM, ASM, PS, ASchw research is carried out within the Collaborative Research Centre 956, sub-projects A6 and C3, funded by the Deutsche Forschungsgemeinschaft (DFG) - project ID 184018867. MP acknowledges funding from the European Unions Horizon 2020 research and innovation program under the Marie Sk\l{}odowska-Curie grant agreement No 664931. This research made use of Astropy,\footnote{http://www.astropy.org} a community-developed core Python package for Astronomy \citep{Astropy-Collaboration:2013aa,Astropy-Collaboration:2018aa}.
\end{acknowledgements}

\bibliographystyle{aa} 
\bibliography{ref}

\begin{appendix}

\moda{
\section{Intensity profiles}
\label{background}
  In Fig.~\ref{f:backgroundcheck} we plot the intensity profiles of the twelve channel maps across the Sgr\, B2(DS) region. The profiles are taken along the RA direction from (17:47:05, $-$28:25:10) to (17:47:38, $-$28:25:10). Since we set \textit{uv} limit to 0.6--50~k$\lambda$, the large scale structures are filtered out, resulting in the background intensity to be $\sim 0$. 
  \begin{figure*}[h]
    \begin{center}
    \begin{tabular}{c}
            \includegraphics[width=0.55\textwidth]{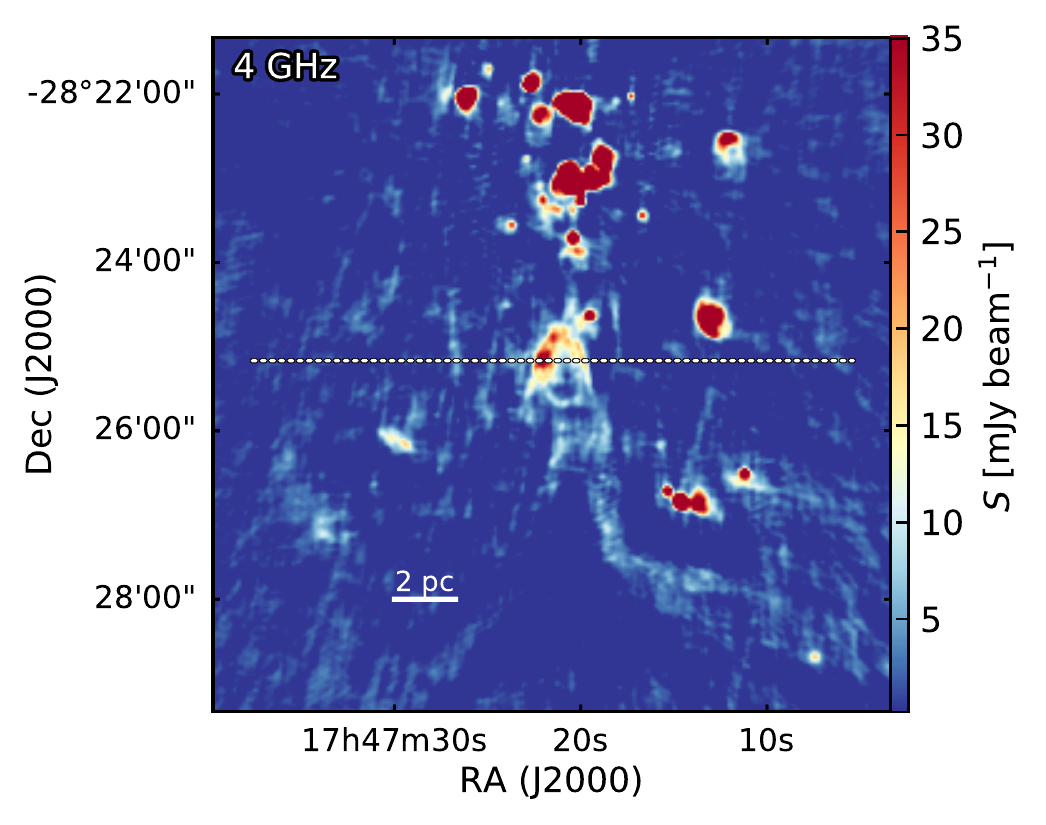} \\   
            \includegraphics[width=0.85\textwidth]{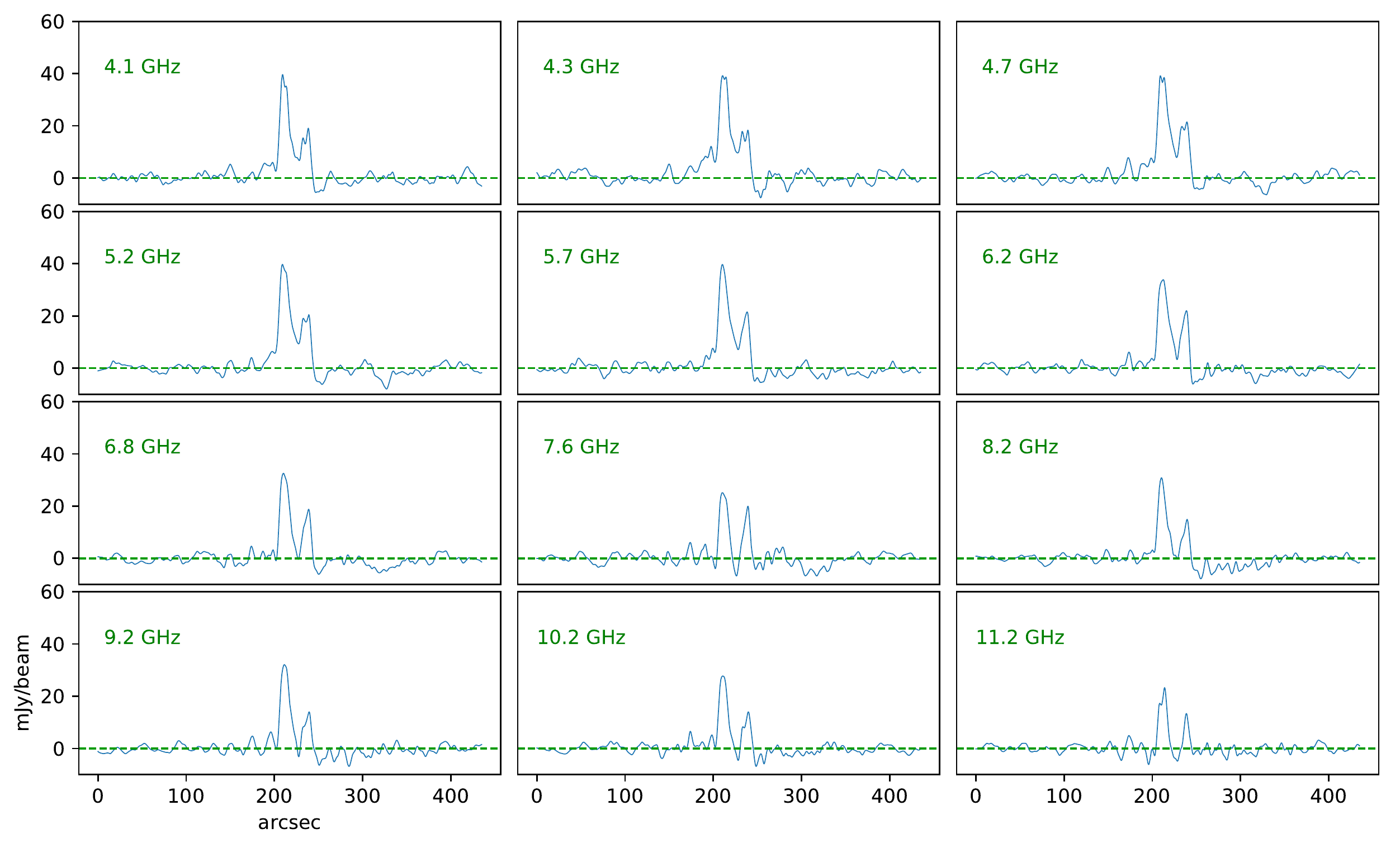}
             \end{tabular}
    \caption{\textit{Top panel:} Continuum emission image  of Sgr\,B2 at 4~GHz. The dashed line indicates the trajectory along which we get the intensity profile. \textit{Bottom panels:} Intensity profiles of the twelve channel maps. The \textit{uv} limit of the twelve images has been  set to 0.6--50~k$\lambda$ and the images have been convolved to a circular beam of 4\arcsec. The dashed line indicates the zero intensity level.}
    \label{f:backgroundcheck}
    \end{center}
  \end{figure*}
}

\section{Combining GMRT and VLA}
\label{appendix_GMRT}

\begin{figure*}[h]
    \begin{center}
    \begin{tabular}{c}
            \includegraphics[width=0.85\textwidth]{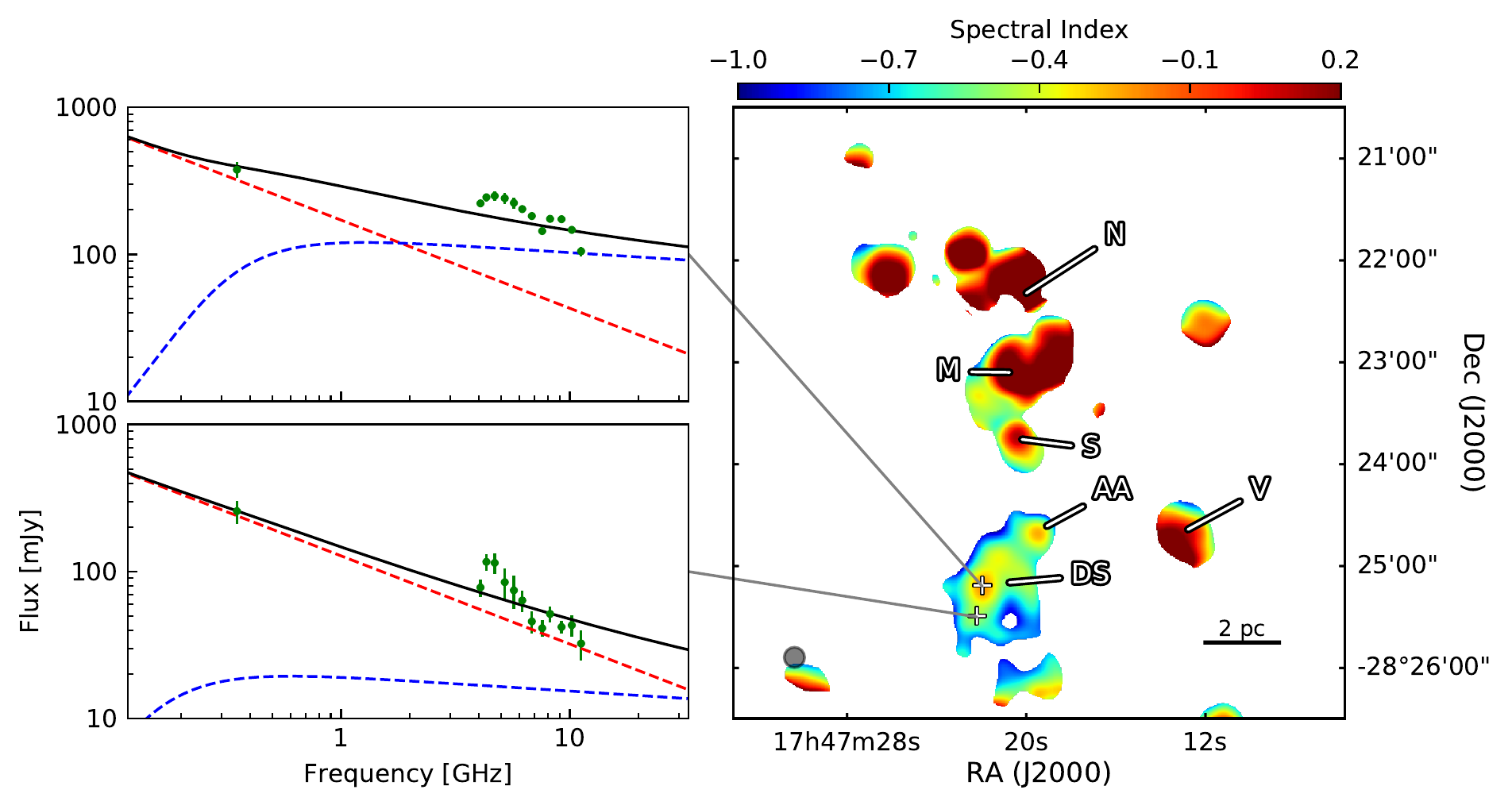}
             \end{tabular}
    \caption{\textit{Right panel:} Spectral index image obtained from the 350~MHz GMRT emission map (Meng et al., in prep.) and the 4~GHz VLA continuum image.  Both images have been obtained after limiting the short $uv$ to 0.6~k$\lambda$, and convolving the images to a common beam of 12.2\arcsec$\times$11.7\arcsec (${\rm PA}=57^\circ$), which corresponds to $\sim 25$~k$\lambda$. \textit{Left panels:} Two examples of SEDs for two selected positions within DS. The observed fluxes are shown with green symbols, all of them measured from images with same spatial filtering and synthesized beam.  For illustration, the thermal (blue dashed curve), non-thermal (red dashed line) and the mixed (black solid curve) components are marked. The curves are just qualitative descriptions of the possible contribution of the thermal and non-thermal components, and do not aim at fitting the data.}
    \label{f:gmrt_mixed_model}
    \end{center}
  \end{figure*}

\modb{The GMRT observations on the Galactic Center are in 350~MHz, with a bandwidth of 32~MHz.  For comparison, we select the GMRT data with \textit{uv} limit 0.6--25~k$\lambda$. The short spacing of 0.6~k$\lambda$ is the same as that of all the 12 VLA channels maps. The beam size of the GMRT image is 12.2\arcsec$\times$11.7\arcsec (${\rm PA}=57^\circ$). We convolved the VLA channels images to the same resolution as the GMRT data, which makes the long \textit{uv} limit of both GMRT and VLA $\sim$ 25~k$\lambda$. We obtain the spectral index in DS by interpolating the flux density between 350~MHz and 12~GHz (see Fig.~\ref{f:gmrt_mixed_model}). The SED of all the 13 channels (see Fig.~\ref{f:gmrt_mixed_model}) shows that the radio emission of DS is a mixture of thermal and non-thermal emission, in consistent with the analysis based on the VLA data. However, the relatively low resolution of the GMRT image makes the study on the small-scale structure of the DS bubble impossible. Therefore, in this paper, we only focus on the VLA images and the sub-pc structures in DS. The full version of the GMRT data will be used to study the large scale structure of the Galactic center region in a separate paper (Meng et al. in prep.).}

\moda{In this section we make use of recently observed GMRT observations to confirm the negative spectral indices measured towards the Sgr\,B2(DS) region. The GMRT observations were taken under the project number (31\_021) and will be presented in detail in a forthcoming publication. In summary, we used the GMRT to observe the Galactic center at 350~MHz, with a total bandwidth of 32~MHz. The observations were centered at the position of Sgr\,B2. For comparison with the VLA data, we have selected data in the \textit{uv} range 0.6--25~k$\lambda$. The short spacing is set to 0.6~k$\lambda$ to match the same scales recovered in the VLA data. The synthesized beam of the GMRT image is $12\farcs2\times11\farcs7$ (${\rm PA}=57^\circ$). We have convolved the VLA images to the same resolution of the GMRT data, which results in the upper limit of the \textit{uv} range for both the GMRT and VLA images to be $\sim$25~k$\lambda$.}

\moda{We obtain the spectral index in DS by interpolating the flux density between the 350~MHz and 12~GHz images. As shown in the right panel of Fig.~\ref{f:gmrt_mixed_model}, the spectral index in DS is negative and in agreement with the spectral index values determined using the VLA data (cf.\ Fig.~\ref{f:spix_err}). In the left panels of Fig.~\ref{f:gmrt_mixed_model} we show the SED of the 13 channel images for two positions in DS. The SEDs in these positions can be describe by a combination of thermal and non-thermal emission, in agreement with the analysis based on the VLA data. However, the relatively coarse angular resolution of the GMRT images makes the study of the small-scale structure of the DS bubble not possible. Therefore, in this paper, we focus on the analysis of the VLA images and the sub-pc structures in DS. A complete analysis of the GMRT data will be used to study the large-scale structure of the Galactic center region in a forthcoming paper (Meng et al., in prep.).}

\section{The RRL maps}
\begin{figure*}[h]
    \begin{center}
    \begin{tabular}{c}
            \includegraphics[width=0.95\textwidth]{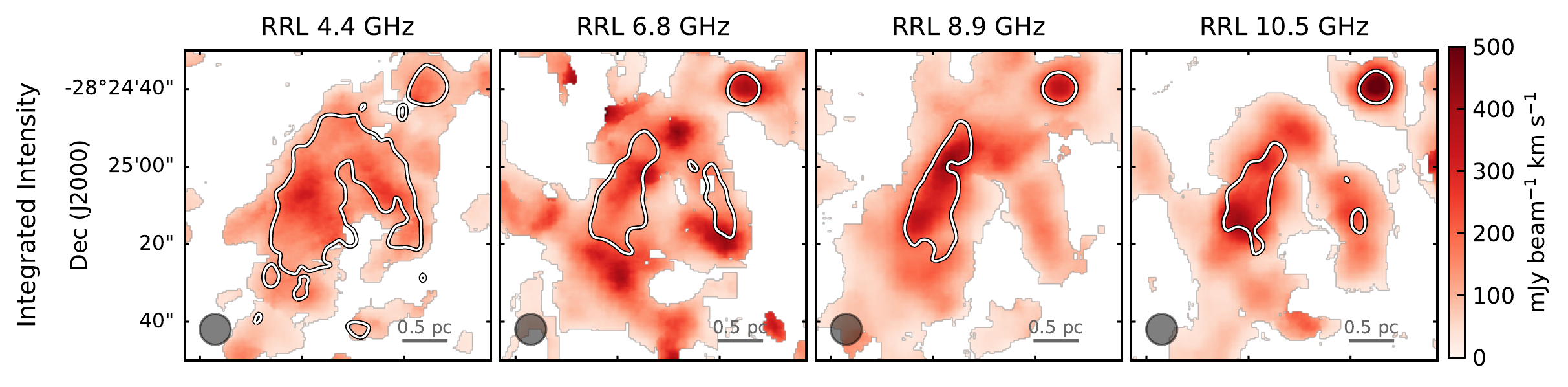} \vspace{-0.2cm} \\
            \includegraphics[width=0.95\textwidth]{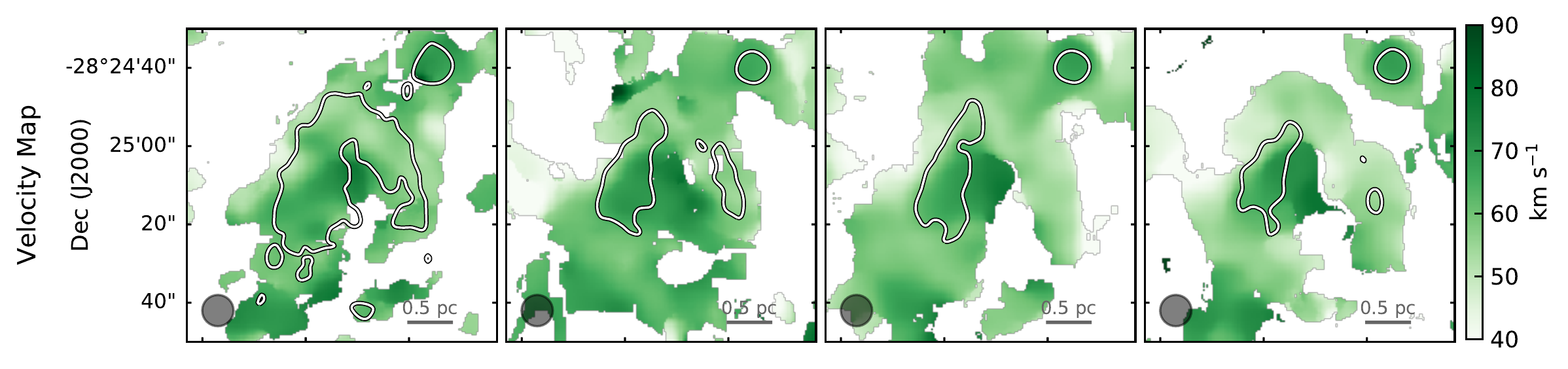} \vspace{-0.2cm}\\
            \includegraphics[width=0.95\textwidth]{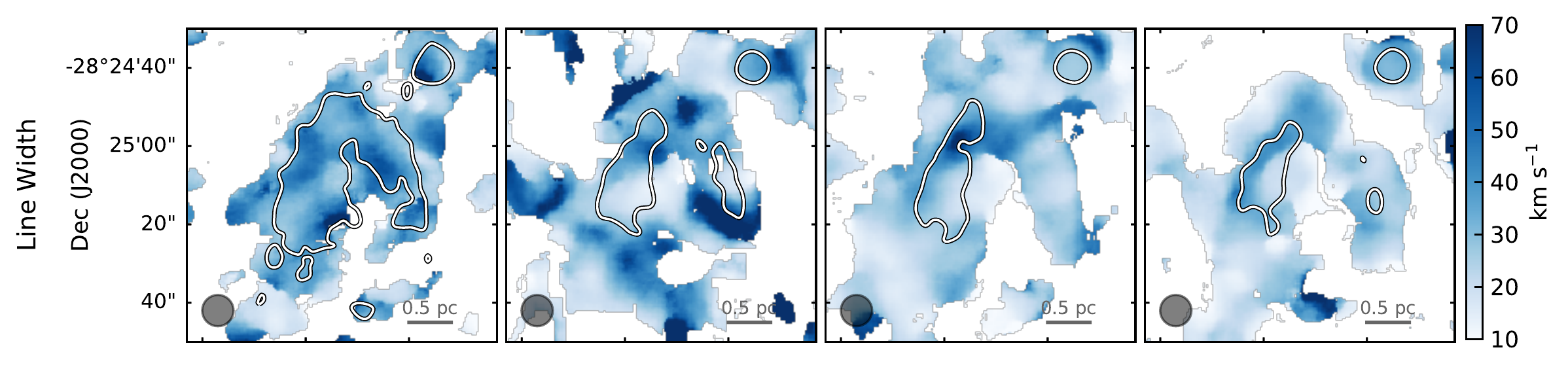} \vspace{-0.2cm}\\
            \includegraphics[width=0.95\textwidth]{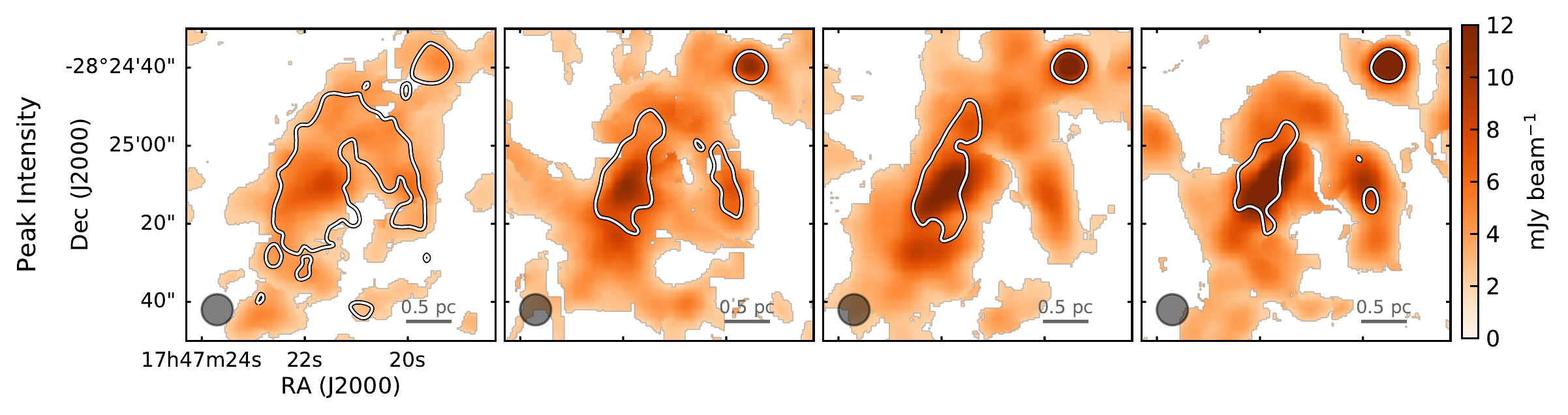} 
    \end{tabular}
    \caption{\moda{Maps} of \moda{the} fit parameters of RRLs in DS. From top to bottom the four rows show \moda{the} integrated intensity, centroid velocity, line width, and peak intensity. The RRLs are fit with \moda{a} Gaussian function. The \moda{synthesized beam} (8 arcsec) is shown at the lower left corner of each panel. Continuum emission at respective frequencies (4.4, 6.8, 8.9 and 10.5~GHz) are overlaid as contours.}
    \label{f:rrl_all}
    \end{center}
  \end{figure*}

  In this section we show the maps of fit parameters of the four stacked RRLs (see Table~\ref{t:rrls}). The fit parameters include integrated intensity, centroid velocity, line width, and peak intensity (see Fig.~\ref{f:rrl_all}).


\end{appendix}
\end{document}